\documentclass[journal,twoside,web]{ieeecolor}
\usepackage{generic}
\usepackage{cite}
\usepackage{amsmath,amssymb,amsfonts}
\usepackage{algorithmic}
\usepackage{graphicx}
\usepackage{textcomp}

\newcommand{\cost}{\Psi} %costs
\newcommand{\price}{\Phi} % prices
\newcommand{\welfare}{\omega} % MM utility

\newcommand{\costpar}{Q}

\usepackage{eurosym}

\usepackage[T1]{fontenc} % Use 8-bit encoding that has 256 glyphs

\usepackage[utf8]{inputenc} % Required for including letters with accents

\usepackage{array,booktabs}
\usepackage{subfig}
\usepackage{makecell}
\usepackage{float}
\usepackage{pdfpages}
\usepackage{subfig} % Required for creating figures with multiple parts (subfigures)

\usepackage{amsmath,amssymb} % For including math equations, theorems, symbols, etc
\usepackage{mathtools}
\usepackage{amsfonts}
\usepackage{bbm}
\usepackage{multirow}
\usepackage{lastpage}
\usepackage{scrtime}
\usepackage{mathrsfs,dsfont} 
\usepackage{eurosym}
\usepackage{amscd}
\usepackage{caption}
\usepackage{varioref} % More descriptive referencing
\usepackage{tikz}
\usetikzlibrary {positioning}
\usepackage{pgfplots}
\pgfplotsset{ % Here we specify options for all figures in the document
	compat=newest, % Which version of pgfplots do we want to use?
	legend style =
	{font=\small \sffamily},
	label style = {font=\small\sffamily},
	every tick label/.append style={font=\small}}
\newcommand{\mc}{\mathcal}
\usepackage[most,skins,theorems]{tcolorbox}
\tcbset{highlight math style={enhanced,
		colframe=red,colback=white,arc=0pt,boxrule=1pt}}
\usetikzlibrary{arrows}

\def\1{\mathds{1}}

\def\R{\mathbb{R}}

\newcommand{\overbar}[1]{\mkern 1.5mu\overline{\mkern-1.5mu#1\mkern-1.5mu}\mkern 1.5mu}

\usetikzlibrary[topaths]
\newcount\mycount
\DeclareMathOperator*{\argmax}{arg\,max}

\DeclareMathOperator*{\diag}{diag}
\DeclareMathOperator*{\rank}{rank}
\newcommand{\ba}{\begin{array}}
\newcommand{\ea}{\end{array}}

\newcommand{\be}{\begin{equation}}
\newcommand{\ee}{\end{equation}}
\newcommand{\se}{\text{ if }}
\newcommand{\ds}{\displaystyle}

\newcommand{\eps}{\varepsilon}

\newcommand{\de}{\mathrm{d}}
\newcommand{\tcr}{\textcolor{red}}
\newcommand{\tcb}{\textcolor{black}}

\newtheorem{theorem}{Theorem}
\newtheorem{lemma}{Lemma}

\newtheorem{corollary}{Corollary}
\newtheorem{example}{Example}
\newtheorem{definition}{Definition}
\newtheorem{remark}{Remark}
\newtheorem{assumption}{Assumption}

\let\labelindent\relax
\usepackage[shortlabels]{enumitem}

\usepackage{comment}
\usepackage[framemethod=tikz]{mdframed}

\usepackage{xparse}

\usetikzlibrary{arrows.meta}

\usepackage{layouts}
\usepackage{url}

\usepackage[framemethod=tikz]{mdframed}

\definecolor{mycolor}{rgb}{0.122, 0.435, 0.698}
\usepackage{graphicx}
\usepackage{hyperref}

\begin{document}    

\title{Equilibria in Network-Constrained Markets with a System Operator}

\author{Giacomo~Como,~\IEEEmembership{Member,~IEEE,}
	Fabio~Fagnani, Leonardo~Massai, and Martina Vanelli,~\IEEEmembership{Member,~IEEE}
	\thanks{A preliminary version of this paper was presented at the 22nd IFAC World Congress \cite{como2023equilibria}.}   
	\thanks{This work was partially supported by the Research Project PRIN 2022 ``Extracting Essential Information and Dynamics from Complex Networks'' (Grant Agreement number 2022MBC2EZ) funded by the Italian Ministry of University and Research.}
%	\thanks{Giacomo Como is with the Department of Mathematical Sciences		“G.L. Lagrange,” Politecnico di Torino, 10129 Torino, Italy, and also with the Department of Automatic Control, Lund University, 22100 Lund,		Sweden (e-mail: giacomo.como@polito.it).}
%	\thanks{Fabio Fagnani is with the Department of Mathematical Sciences		“G.L. Lagrange,” Politecnico di Torino, 10129 Torino, Italy (e-mail:		fabio.fagnani@polito.it).}
	\thanks{
G.~Como and F.~Fagnani are with the Department of Mathematical Sciences ``G.L.~Lagrange'', Politecnico di Torino, 10129 Torino, Italy (\texttt{\{giacomo.como,fabio.fagnani\}@polito.it}). G.~Como is also with the Department of Automatic Control, Lund University, SE-22100 Lund, Sweden.}
	\thanks{L.~Massai is with the Institute of Mechanical Engineering, \'Ecole Polytechnique Fédérale de Lausanne (EPFL), CH-1015 Lausanne, Switzerland (e-mail: \texttt{l.massai@epfl.ch}).}
	\thanks{\tcb{M.~Vanelli is with the Engineering and Technology Institute Groningen, University of Groningen, 9747AG, The Netherlands (e-mail: \texttt{m.vanelli@rug.nl})}.}
	%\thanks{\tcr{M.~Vanelli is with the ICTEAM Institute Université catholique de Louvain, 1348 Ottignies-Louvain-la-Neuve, Belgium (e-mail: martina.vanelli@uclouvain.be)}.}
}       
\maketitle
\thispagestyle{empty}
%   \thanks{We thank Sergio Augusto Angelini for his help in developing the code of the case study. }
%%%% template IFAC %%%%%%
%\begin{frontmatter}

%\title{Equilibria in Network Constrained Markets with a system operator\thanksref{footnoteinfo}} 

%\thanks[footnoteinfo]{This work was supported by Ministero dell'Istruzione, dell'Universita e della Ricerca [Grant 	E11G18000350001 and Research Project PRIN 2017 “Advanced Network Control of Future Smart	Grids”] and the Compagnia di San Paolo.}

%\author[First]{Giacomo Como} 
%\author[First]{Fabio Fagnani} 
%\author[Second]{Leonardo Massai}
%\address[First]{Department of Mathematical Sciences ``G.L.~Lagrange'', Politecnico di Torino, Corso Duca degli Abruzzi 24, 10129 Torino, Italy\\ 	(e-mail: \{giacomo.como,fabio.fagnani\}@polito.it).}
% \address[Second]{Ecole Polytechnique Fédérale de Lausanne (EproofL) Institute of Mechanical Engineering CH-1015 Lausanne, Switzerland\\ 	(e-mail: l.massai@eproofl.ch).}

\begin{abstract}                % Abstract of not more than 250 words.
We study a networked economic system composed of $n$ producers supplying a single homogeneous good to a number of geographically separated markets and a centralized authority, called the system operator (SO). Producers compete {\it \`a la} Cournot by choosing the quantities of the good to supply to each market they have access to in order to maximize their profit. Every market is characterized by an inverse demand function returning the unit price of the considered good as a function of the total available quantity. Markets are interconnected by a dispatch network through which quantities of the considered good can flow within finite capacity constraints and possibly satisfying additional linear physical constraints. Such flows are determined by the action of a SO, who aims at maximizing a designated welfare function. 

We model such competition as a strategic game with $n+1$ players: the producers and the SO. We first establish the existence of pure-strategy Nash equilibria under standard concavity assumptions. Then, we identify sufficient conditions for the game to be exact potential with an essentially unique Nash equilibrium. Next, we present a general result connecting the optimal action of the SO with the network capacity constraints. 
For the commonly used Walrasian welfare, our findings establish a connection between capacity bottlenecks in the market network and the emergence of price differences between markets separated by saturated lines. This phenomenon is frequently observed in real-world scenarios, for instance in power networks. Finally, we validate the model with data from the Italian day-ahead electricity market.\end{abstract}


\section{Introduction}
The study of network effects in modern marketplaces  has attracted significant attention in recent years \cite{easley2010networks}. In particular, a growing body of literature underscores the limitations of classical competition models, which typically depict multiple producers operating within a single, isolated market. These basic models often overlook the increasing interconnectedness that characterizes power systems \cite{Neuhoff2005}, transportation and infrastructure networks, global supply chains and the Internet, and they fail to describe emergent features of such networked systems. Consequently, several works in the literature are devoted to extend classical oligopoly models \cite{vives1999oligopoly}, such as Bertrand, Cournot, and Stackelberg competitions, to include network structures that reflect real-world market interdependencies. 

Our paper specifically contributes to the expanding literature on networked Cournot competition. Originally introduced in 1838 \cite{cournot1838recherches}, the Cournot competition  is a fundamental economic model wherein firms compete in a single market by selling a homogeneous good. In this model, each firm sets the amount of production in order to maximize its profit. The effect of networks in Cournot competitions was first studied in \cite{bulow1985multimarket} and subsequently in \cite{Abolhassani.ea:2014} and \cite{Bimpikis2019}.  In these works, the classical model of Cournot competition is extended by considering multiple firms operating in different markets: producers and markets are coupled via a bipartite graph, where links indicate the markets each producer can supply to 
and producers' cost functions are typically non-separable across the markets they participate in.

Our work differs from the classical setting and in particular takes inspiration from  \cite{Cai2019}, where a networked Cournot competition among multiple producers is studied together with the presence of an additional player there referred to as \emph{market maker}: this centralized authority moves supply between geographically separated markets via the constrained network to achieve a desirable state of the system.  This setting is particularly relevant in energy markets \cite{Neuhoff2005}, where a centralized authority typically solves a dispatch problem by leveraging the offers and bids from generators and retailers, with the objective of maximizing a specific metric of social welfare, under the operational constraints of the grid. This is precisely the model we aim at capturing by introducing a \emph{system operator}\footnote{Our choice of the term "system operator" rather than ``market maker'' is to better align with the literature of equilibria with network transmission constraints \cite{downward2010cournot}.} (SO) that is in charge of matching the demand and supply of goods and of their transportation between different markets. The SO acts as an independent regulated entity and, through the design of its utility function, it mitigates the potential exercise of market power by producers.  

In this paper, we consider a constrained dispatch network modeled by a graph where nodes indicate the markets and links represent the finite-capacity physical lines connecting them. Producers engage in a networked Cournot game within this framework alongside a SO that can move the good among different markets along the available links, while striving to maximize a specific welfare function. In contrast to \cite{Cai2019}, in our study firms are allowed to sell in multiple markets and a more general class of price, cost, and welfare functions is considered. Additionally, the focus of our analysis is not on the design of the SO's utility function, but it is rather directed to studying the Nash equilibria of the game, particularly the influence that the network and the capacity constraints have on these equilibria.
	
Our main contributions are the following. We first establish the existence of pure-strategy Nash equilibria under standard concavity assumptions. In the special case when the SO's utility function is the Walrasian welfare (c.f.~\cite{Johari2005} and \cite{Cai2019}) and the price functions are affine, we show that the considered game is potential. Moreover, the specific form of the potential function guarantees that in this case the pure-strategy Nash equilibrium is essentially unique and can be efficiently computed by solving a concave optimization problem.  Second, we characterize the general structure of the optimal flows chosen by the SO and analyze the restrictions posed on such flows by the capacity constraints within the market network. Notably, these results hold quite in general, when the system is not necessarily in a Nash equilibrium, as they only require the SO to be selecting an optimal flow. Restricting to the case where the SO's utility function corresponds to the Walrasian welfare, we establish a key finding for optimal flows: if the SO is playing a best response and prices differ across the endpoints of a link, then	some physically related link (not necessarily the one across which the price difference is observed) is congested. We further make this connection explicit in terms of the graph-theoretic properties of the network and the physical flow model. This result formally connects price discrepancies with capacity constraints—a phenomenon frequently observed in real power networks \cite{who}. We further provide a sufficient condition to ensure non-negativity of the demand in each market without requiring an explicit constraint to enforce this (and thus avoiding the use of Generalized Nash equilibria).  Finally, we validate our model with real data from the Italian day-ahead electricity market.

\subsubsection*{Related work}
Following insights from \cite{bulow1985multimarket}, Network Cournot Competitions (NCC) were first introduced in \cite{Abolhassani.ea:2014}, which considers bipartite graph models in which firms are on one side and markets on the other side, with links indicating whether a firm has access to a market or not. The authors prove that the NCC with linear price functions is an ordinal potential game and the potential function is concave provided that the cost functions of the firms are convex. In \cite{Bimpikis2019}, the NCC is studied in a simpler setting with linear inverse demand functions and quadratic production costs. The authors explicitly characterize how equilibrium quantities depend on the competition structure and explore the welfare implications of entry, mergers, or changes in the environment’s primitives. 

A challenge in Cournot modeling of power markets has been the inclusion of transmission constraints, which are widely acknowledged to be a major reason why suppliers can manipulate price. One approach, introduced in \cite{borenstein1997competitive} for symmetric duopolies and studied in \cite{downward2010cournot} for radial networks, involves a centralized dispatch: firms compete {\it \`a la} Cournot, and then an economic dispatch problem is solved to reallocate energy across markets while satisfying network constraints. Models of competitions in networks were also considered for Supply Function Equilibria in \cite{wilson2008supply} and \cite{Holmberg.Philpott:2018}, which consider a transport-constrained network with local demand shocks where spatially distributed oligopoly producers compete with supply functions, as in wholesale electricity markets. In such models, however, equilibria may not exist since a producer in an importing node can find it profitable to deviate from a locally optimal profit maximum by withholding production in order to congest imports and push up the nodal price.

	In other studies where centralized dispatch is not included, such as \cite{oren1997economic}, the satisfiability of physical capacity bounds is guaranteed by introducing explicit constraints in the configuration space of the producers, thus leading to Generalized Nash Equilibria (GNE). This approach has been criticized, e.g., in \cite{stoft1999financial}  for using a non-standard equilibrium concept in which producers behave as if contributing to congestion were prohibited, thereby replacing true non-cooperative interaction with an exogenous rule. The work	\cite{willems2002modeling} defended the GNE approach by modeling the intervention of a market operator who reduces quantities when capacities are exceeded. The key difference between the centralized approach and that in \cite{willems2002modeling} lies in how the SO intervenes: in one case, offered quantities are redistributed without altering production, while in the other, quantities themselves are adjusted based on rules such as “all-or-nothing,” proportional allocation, or efficiency-based criteria.  %Due to the shared constraints, computing a GNE is usually hard and distributed algorithms have been developed to efficiently compute equilibrium solutions \cite{ zhu2016distr}. %In \cite{peng2019generalized}, an operator splitting approach is proposed to compute GNE over network systems via a fully distributed algorithm, where each player relies only on local information and neighbor communication, but must observe the decision variables involved in its local objective function to evaluate its local gradient. The partial-information setting has been studied in \cite{2020p}, and several other scenarios have been considered, e.g., time-varying networks \cite{9130079} and stochastic costs \cite{franci2020distributed}.
	
	Our approach is closely related to the centralized economic dispatch framework but introduces a key modification: the market operator participates simultaneously as a strategic player alongside firms. This contrasts with the Stackelberg game formulation adopted, e.g., in \cite{borenstein1997competitive}, \cite{downward2010cournot}, and \cite{xu2017onthe}, where the producers act as leaders and the SO acts as a follower.  Prior work has explored this idea \cite{Cai2019}, demonstrating that equilibrium outcomes are strongly influenced by the market operator’s objective function and providing conditions under which the resulting game is a potential game. The simultaneous approach has the advantage of ensuring existence of Nash equilibria, facilitating deeper theoretical analysis.

The goal of the SO is to maximize social welfare. Social welfare in competition models has been explored in several works, such as \cite{Johari2005}, \cite{johari2005efficiency}, which shed light on the robustness of the pricing mechanism in network resource allocation when users behave selfishly and anticipate the effects of their actions on prices. The paper \cite{johari2004efficiency}  delves into the concept of the price of anarchy in congestion games, examining how selfish behavior affects system efficiency, while \cite{tsitsiklis2013profit} examines the classical Cournot oligopoly model and investigates the potential gain in profits if the oligopolists were to collude, or conversely, the reduction in profits due to competition. Finally, \cite{zhang2019competition} studies the impact of coalition formation on the efficiency of Cournot games in the presence of uncertainties. 

Other works have focused on specific applications, particularly electricity power models where the physical network connecting different markets plays a crucial role \cite{conteras2004num}.  In \cite{Barquin8}, the authors analyze a constrained power network that connects various markets and producers under a Cournot competition framework. The authors develop an iterative algorithm to find the Nash equilibrium, which takes into account how the production at a specific node affects the whole network, consequently explaining the opportunities for the producers to exercise market power. In \cite{Neuhoff2005} a numerical analysis is presented regarding the sensitivity of Nash equilibria within a networked Cournot competition in a transmission-constrained electricity market, revealing that Cournot equilibria are indeed highly sensitive to assumptions regarding market design. A two-settlement electricity market incorporating both forward and spot markets is introduced in \cite{Yao2008}, which accounts for flow congestion, demand uncertainty, system contingencies, and market power. The model assumes linear demand functions, quadratic cost functions, and a lossless DC power network. 

We remark that, in energy markets, producers typically submit bid pairs of quantity and price, rather than deciding quantities as in Cournot models. This is commonly captured via supply function equilibria \cite{klemperer1989supply}. However, the analysis of such models is significantly more complex in the presence of network constraints. In this respect, \cite{willems2009cournot} shows that, given their similar explanatory power, Cournot models are  well suited for the study of market rules or congestion allocation mechanisms.

Finally, we acknowledge a closely related line of work on routing games with capacity constraints such as \cite{correa2004selfish} and \cite{ gourves2015congestion}. These studies analyze equilibrium properties in congestion games with link or node capacities. While our setting differs from these ones, some of the resulting insights are closely related. In particular, capacity constraints may induce equilibrium utility differences, a phenomenon that parallels the price differences caused by saturated links in our model. 

\subsubsection*{Structure of the paper and notation}
The rest of this paper is organized as follows. The remainder of this section is devoted to the introduction of some notational conventions used throughout. In Section \ref{sec2}, we present the model of networked Cournot competition with a system operator (SO). In Section \ref{sec3}, we prove existence of Nash equilibria under standard assumptions of convexity of the production cost functions and concavity of the inverse demand and welfare functions (see Assumption \ref{assumption:cost+price+welfare}). We further prove that, when all the market inverse demand functions and the Walrasian welfare function are affine, the game is potential with a\tcb{n essentially} unique Nash equilibrium.  In Section \ref{sec4}, we prove for the Walrasian welfare that the emergence of price differences is linked with capacity bottlenecks and critical links in the distribution network. We also derive sufficient conditions for net consumptions to be positive at Nash equilibrium and compare our model with the sequential Stackelberg–Cournot framework considered in \cite{borenstein1997competitive, downward2010cournot}. In Section \ref{sec5}, we present a case study featuring  the Italian day-ahead market. Finally, in Section \ref{sec6} we draw some conclusions and discuss current and future research. 

Throughout the paper, we denote vectors in lower case, matrices in upper case,
and sets with calligraphic letters. We indicate with $\1$ the all-1 vector and with $I$ the identity matrix, regardless of their dimension. The transpose of a matrix $A$ in $\R^{m\times n}$ is denoted by $A'$ in $\R^{n\times m}$.  A subscript associated to vectors, for instance, $v_{\mc U}$, represents the sub-vector that is the restriction of the vector $v$ in $\R^n$ to the set of indices $\mc U \subseteq \{1, \dots, n   \}$. For a square matrix $A$, $\operatorname{diag}(A)$ denotes the vector containing the diagonal elements of $A$. For any $a,b \in \mathbb{R}$, we denote with $a \wedge b := \min\{a,b\}$ and $a \vee b := \max\{a,b\}$ the minimum and maximum of $a$ and $b$, respectively.

\section{Model and problem statement} \label{sec2}
\subsection{Market Network} 
We consider a nonempty finite set  $\mc N = \{1, \dots, n\}$ of competing producers, each one selling some quantity of the same homogeneous good on a nonempty subset of a finite set of markets $\mc M = \{1, \dots, m\}$. We assume that the markets are interconnected by a finite set of links $\mc L=\{1,\ldots,l\}$. Every link $k$ in $\mc L$ is to be interpreted as a physical connection that can carry flow of the considered good between its end-nodes. We assume that every link $k$ in $\mc L$ has been oriented\footnote{The choice of such orientation is arbitrary and does not affect any results.} from its tail node $\sigma_k$ in $\mc M$ to its head node $\tau_k$ in $\mc M\setminus\{\sigma_k\}$  so that  a positive flow $f_k>0$ is to be understood as moving a quantity $f_k$ of the good from market $\sigma_k$ to market $\tau_k$, whereas a negative flow $f_k<0$ is to be understood as moving a quantity $-f_k$ of the good in the opposite direction, i.e., from market $\tau_k$ to market $\sigma_k$. Importantly, such flows are limited by capacity constraints 
\be\label{capacity-constraints}-c_k^-\le f_k\le c_k^+\,,\qquad \forall k\in\mc L\,,\ee
where $\tcb{c_k^+}\ge0$ and $c_k^-\ge0$ are the maximum flow capacities of link $k$ in the direct and reverse direction, respectively. We stack all the link flows and capacity values in vectors $f$ \tcb{in $\R^l$, and} $c^+$\tcb{,} $c^-$ in $\R_+^l$. Notice that we allow for the possibility of parallel links between markets (i.e., distinct links with both the same tail node and the same head node), but not for self-loops, i.e., links whose tail node coincides with their head node. In general, for a link $k$, the capacities $c^+_k$ and $c_k^-$ may differ.
In particular, we will refer to links $k$ in $\mc L$ such that either $c_k^+>0$ and $c_k^-=0$ or \tcb{$c_k^->0$} and $c_k^+=0$ as \emph{directed}. 

The interconnection structure between producers and markets and among the latter is completely described by the producer-market adjacency matrix\footnote{Notice that we allow producers to possibly sell on multiple markets. The setting studied in \cite{Cai2019} ---where every producer can sell on a single market--- can be recovered as a special case in our framework, with $\sum_{j \in \mc M}A_{ij}=1$ for every $i$ in $\mc N$.} $A$ in $\{0,1\}^{n\times m}$, whose entries are defined by
\be\label{Aij} A_{ij}=\left\{\ba{ll}1&\se\text{producer }i\text{ can sell on market }j\\0&\se \text{producer }i\text{ cannot sell on market }j\,,\ea\right.\ee 
and the market-link incidence matrix $B$ in $\{0,\pm1\}^{m \times l}$,  whose entries are defined by
\be\label{Bij} B_{jk}=\left\{\ba{ll}1&\se \tau_k=j\\ -1&\se \sigma_k=j\\0&\se \sigma_k\ne j\ne \tau_k \,,\ea\right.\ee 
respectively. For a positive integer $\ell\ge1$, a length-$\ell$ path between markets $j$ and $j^*$ in $\mc M$ is an $\ell$-tuple of links $(k_1,k_2,\ldots,k_l)$ for which there exists a $(\ell+1)$-tuple of markets $(j_0,j_1,\ldots,j_{\ell})$ such that: $j_0=j$, $j_{\ell}=j^*$; for every $1\le h\le\ell$, link $k_h$ connects markets $j_{h-1}$ and $j_h$, i.e., $\{\sigma_{k_h},\tau_{k_h}\}=\{j_{h-1},j_h\}$; and $j_{g}\ne j_{h}$ for every $0\le g<h\le\ell$, except for possibly $j_0=j_\ell$ in which case the path is called a cycle. Throughout,  we assume that the market network is weakly connected, i.e., there exists a path between every two markets $i\ne j$ in $\mc M$. This is equivalent \tcb{to} $\rank(B)=m-1$.

For every flow vector $f$ in $\R^l$, the entries $r_j$ of the vector \be\label{r=Bf}r=Bf\,,\ee in $\R^m$ represent the net quantities of the good moved by the SO to (if $r_j>0$) or from (if $r_j<0$) market $j$ in $\mc M$. Observe that, since the columns of the market-link incidence matrix $B$ are zero-sum, we have that $\sum_{j\in\mc M}r_j=\1'Bf=0$, i.e., $r$ is a zero-sum vector, corresponding to the fact that the SO neither produces nor consumes the commodity but only moves it among the markets. As it will be clear later, $r_i$ will coincide with the difference between the consumption and the production in market $i$, hence \eqref{r=Bf} is to be interpreted as \tcb{Kirchhoff}'s first law.

We then consider a matrix $H$ in $\R^{l\times l}$ satisfying the identity 
\be\label{BH=B}BH=B\,.\ee
Such a matrix $H$ maps vectors $y$ in $\R^l$ ---to be interpreted as decision variables of the SO, as illustrated in detail later--- into physical flow vectors $f=Hy$. Equation \eqref{BH=B} implies \tcb{that} the net quantities of the good that the physical flow vector $f=Hy$ moves among the different markets are given by $r = Bf = BHy = By$.\footnote{In fact, one could have equivalently parametrized the decision variables of the SO by $r=By=Bf$. However, we prefer working with the parametrization in $y$ as the capacity constraints are more explicit in this formulation.} We refer to the tuple \be\label{network}(A,B,c^-,c^+,H)\,,\ee with $A$ in $\{0,1\}^{n\times m}$,  $B$ in $\{0, \pm 1\}^{m \times l}$, $c^-$ and $c^+$ in $\R_+^l$, and $H$ in $\R^{l\times l}$ satisfying \eqref{BH=B}  as the \emph{market network}. When the market network contains no cycles, we necessarily have that $H=I$ is the identity matrix. Indeed, in this case, since the market network is weakly connected, we have that $l=m-1=\rank(B)$, so that the matrix $B'B$ is invertible and \eqref{BH=B} implies that 
$H=(B'B)^{-1}B'BH=(B'B)^{-1}B'B=I\,.$

On the other hand, for general network topologies (containing cycles), there are several possible matrices $H$ satisfying \eqref{BH=B}. In particular, our framework covers two main examples of matrix $H$ as special cases. The first one is when $H=I$ is simply the identity matrix: in this case, the physical flow $f=y$ is directly designed by the SO within the link capacity constraints. In this case, the flow $f=y$ satisfies \tcb{Kirchhoff}'s first law \eqref{r=Bf}, but not, in general, \tcb{Kirchhoff}'s second law. 

%The second special case covered in our framework is that of a power network  where the flows arrange themselves according to Kirchhoff's laws for DC load flow. This is usually modeled using a voltage phase angle $\theta_j$ at each node $j$ and a positive line reactance $u_k>0$  \cite{downward2010cournot}. The flow on every link $k$ in $\mc L$ must then satisfy 	$u_k f_k= \theta_{\tau_k} - \theta_{\sigma_k}\,,$ which can be written compactly as \be\label{eq:second_law} f=DB'\theta\,,\ee where $D$ in $\R^{\mc L\times\mc L}$ is a diagonal matrix with diagonal entries $D_{kk}=1/{u_k}$ for every $k$ in $\mc L$. Nodal power injections are given by $r=Bf=BDB'\theta=L\theta$, where $L=BDB'$ is the combinatorial Laplacian matrix. 		Let then $L^{\dagger}$ denote the Moore-Penrose pseudo-inverse of $L$, and define the matrix \be\label{H-electric} H=DB'L^\dagger B\,.\ee	

%\tcb{The second special case corresponds to a power network operated under the DC load-flow approximation \cite{downward2010cournot}. 
The second special case covered in our framework is that of a power network  where the flows arrange themselves according to Kirchhoff's laws for DC load flow\tcb{\cite{downward2010cournot}. Let $\theta$ in $\mathbb{R}^m$ denote the vector of voltage phase angles and let $D$ in $\mathbb{R}^{l\times l}$ be the diagonal matrix of line susceptances, with $D_{kk}=1/u_k$ for each line $k$ in $\mathcal L$, where $u_k>0$ is the line reactance. The DC power-flow equations are
\be\label{eq:second_law}
f = DB' \theta,
\qquad
r = Bf = BDB'\theta = L\theta,
\ee
where $L=BDB'$ is the weighted Laplacian matrix.}		Let then $L^{\dagger}$ denote the Moore-Penrose pseudo-inverse of $L$, and define \be\label{H-electric} H=DB'L^\dagger B\,.\ee	Since
	$\operatorname{Im}(L)=\operatorname{Im}(B)=\{r\in\R^m:\sum_jr_j=0\}$ because the network is weakly connected, we have $BH = BDB' L^\dagger B = LL^\dagger B = B,$ and therefore identity \eqref{BH=B} is satisfied when $H$ is defined by \eqref{H-electric}. Observe that, in this case, for every $y$ in $\R^l$, the physical flow $f = Hy$ satisfies \eqref{eq:second_law} with $\theta = L^\dagger By$: this implies that the rescaled physical flow $D^{-1}f$ sums up to $0$ on every cycle, i.e.,  Kirchhoff's second law holds true.

\begin{figure}\begin{center}\includegraphics[width=4cm]{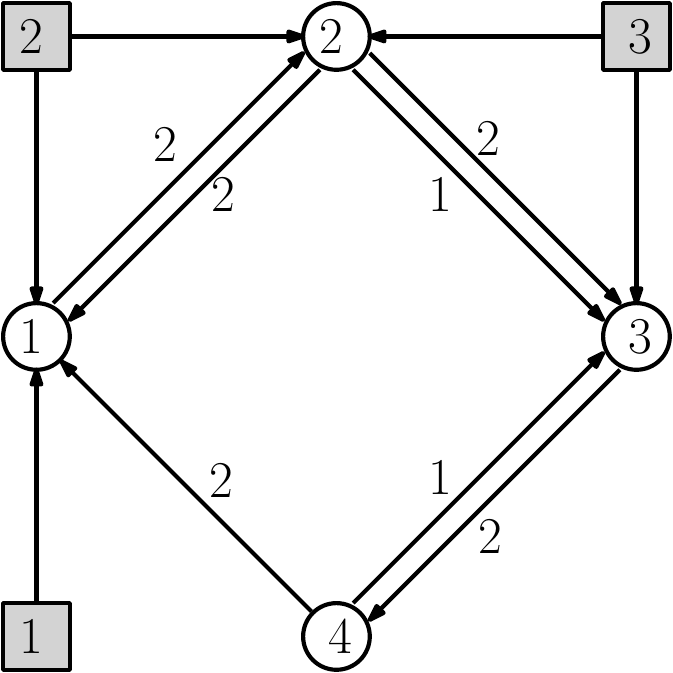}\end{center}\caption{\label{fig:ex-1}The market network of Example \ref{example:1}. Square nodes represent producers, circle nodes represent markets. The labels on the links represent their capacities in the given direction. }\end{figure}
\begin{example}\label{example:1}
	The market network \eqref{network} displayed in Figure \ref{fig:ex-1}  has $n=3$ producers (represented as square nodes), $m=4$ markets (represented as circle nodes), $l=5$ links, producer-market adjacency matrix and 	market-link incidence matrix 
	$$A=\!\left(\!\ba{cccc}1&0&0&0\\1&1&0&0\\0&1&1&0\ea\!\right),\ B=\!\left(\!\ba{ccccccc}1&-1&0&0&0\\0&1&-1&-1&0\\0&0&1&1&-1\\-1&0&0&0&1 \ea\!\right),$$ 
respectively, and  capacity vectors $c^+=(2,2,1,2,2)$ and  $c^-=(0,2,0,0,1)$.
Link $1$ is directed from market $4$ to market $1$ and has capacity $c^+_1=2$, while link $2$ is \tcb{undirected} as it connects markets $1$ and $2$ with capacity $\tcb{c^+_2}=c_2^-=2$ in both directions. On the other hand, links $3$ and $4$ are parallel and directed from market $2$ to market $3$, with capacities $c^+_3=1$ and $c_4^+=2$, respectively. 
Finally, link $5$ can carry flow between markets $3$ and $4$ in both directions with different capacities: $c_5^-=1$ from market $4$ to market $3$ and $c^+_5=2$ from market $3$ to market $4$. 

Observe that this network contains cycles $(1,2,3,5)$, $(1,2,4,5)$, and $(3,4)$. In fact, there are several matrices $H$ in $\R^{l\times l}$ satisfying \eqref{BH=B}, besides the identity matrix $I$, in particular the ones defined as in \eqref{H-electric}.  E.g., if the  reactances are $u_1=1$, $u_2=u_3=1/2$, $u_4=1/3$, and $u_5=1/4$, so that the combinatorial Laplacian matrix is  $$ L=\left(\ba{cccc}5&-1&0&-4\\-1&5&-4&0\\0&-4&7&-3\\-4&0&-3&7\ea\right)\,,$$ then formula \eqref{H-electric} gives us \be\label{H=4}H=\frac1{44}\left(\ba{ccccc}38&-24&-6&-6&-8\\-6&20&-6&-6&-8\\-3&-12&19&19&-4\\-3&-12&19&19&-4\\-6&-24&-6&-6&36\ea\right)\,,\ee which satisfies condition \eqref{BH=B}. \tcb{%Direct inspection of \eqref{H=4} shows that rows $3$ and $4$ of $H$ coincide and that $(3,\,12,\,6,\,0,\,4)\,H=0$; since 
Observe that $\rank(H)=\rank(B)=3$, and that the 
range of $H$ consists exactly of the vectors $f$ in $\R^l$ such that
\be\label{eq:rangeH}
f_3=f_4\,,\qquad 3f_1+12f_2+6f_3+4f_5=0\,.
\ee
In fact, the above are nothing but Kirchhoff's second law on the two cycles $(3,4)$ and $(1,2,3,5)$, respectively.}
\end{example}

\subsection{Producers and system operator}

We model producers as strategic competitors who choose the quantities of the good to sell on the markets available to them, with the objective of maximizing their profits. Specifically, the quantity of the good that a producer $i$ in $\mc N$ sells on a market $j$ in $\mc M$ that it has access to (i.e.,  such that $A_{ij}=1$) is  denoted by $x_{ij}\ge0$. For mere notational convenience, we also introduce the variable $x_{ij}=0$ for every  producer $i$ in $\mc N$ and market $j$ in $\mc M$ not accessible to $i$ (i.e., such that $A_{ij}=0$). All these quantities are assembled in a vector  $x_i=(x_{ij})_{j\in\mc M}$ that represents the action of producer $i$. The action set of producer $i$, to which all such vectors $x_i$ belong, is defined as \be\label{Ai-def}\mc A_i=\left\{v\in\R_+^{m}:\,A_{ij}=0\Rightarrow v_j=0\right\}\,.\ee  Furthermore, we denote by $x=(x_{ij})_{i\in\mc N,j\in\mc M}$ in  $$\mc X=\prod_{i\in\mc N}\mc A_i\subseteq\R_+^{n\times m}\,,$$ the matrix collecting all the quantities sold by the different producers in the different markets. For a producer $i$ in $\mc N$, we also use the notation $$x_{-i}=(x_{hj})_{h\in\mc N\setminus\{i\},j\in\mc M}\,,$$  to indicate the matrix in  $\mc X_{-i}=\prod_{\tcb{h}\in\mc N\setminus\{\tcb{i}\}}\mc A_{\tcb{h}}\subseteq\R_+^{(n-1)\times m}\,,$ obtained by removing the $i$-th row from $x$: such matrix $x_{-i}$ collects the actions of all producers but $i$. 
	 
A SO joins the producers as an additional player whose action is a vector $y$ in $\R^l$ inducing a physical flow vector $f=Hy$ that satisfies the capacity constraints \eqref{capacity-constraints}. As such, the action space of the SO is the set  \be\label{Y-def}\mc Y=\left\{y\in\R^l:\, -c^-\le H y\le c^+\right\}\,.\ee 
 
The quantities of the good $x$ sold by the producers and the flow $f=Hy$ induced by the action of the SO jointly determine the total net consumption \be\label{zj}z_j=\sum_{i\in\mc N}A_{ij}x_{ij}+\sum_{k\in\mc L}B_{jk}f_k=\sum_{i\in\mc N}A_{ij}x_{ij}+\sum_{k\in\mc L}B_{jk}y_k\,,\ee in every market $j$ in $\mc M$, where the last identity follows from \eqref{BH=B}. Equation \eqref{zj}  can be more compactly rewritten as \be\label{z}z=\diag(A'x)+Bf=\diag(A'x)+By\,,\ee where $z=(z_j)_{j\in\mc M}$ is the vector of the market's net consumptions. 

\begin{remark}\label{remark:negative-demand} While the first addend in the right-hand side of \eqref{zj} is always nonnegative, the second addend may be negative. In fact, we are allowing the quantity $z_j$ defined in \eqref{zj} to possibly take also negative values. In equilibrium, this may lead to certain nodes exhibiting negative (non-strategic) demand that is not associated with any producers. Such outcomes may be viewed as a modeling artifact. To address this limitation, in Section \ref{sec:positive}, we identify sufficient conditions under which $z_j$ is nonnegative for every market $j$ in $\mathcal{M}$ in equilibrium.
		\end{remark}\medskip

We assume that every producer $i$ in $\mc N$ incurs a production cost $\cost_i(x_i)$ when choosing to sell  a vector $x_i$ of quantities in the different markets and that the unit price for the good sold on a market $j$ in $\mc M$ is a function $\price_{j}\left(z_{j}\right)$ of the total net consumption in such market. The net profit of every producer $i$ in $\mc N$ is given by the difference between its revenue $\sum_{j\in\mc M}A_{i j}x_{ij}  \price_{j}(z_j)$ and its production cost $\cost_i(x_i)$. On the other hand, the SO %models a SO aiming 
\tcb{aims} at maximizing the total welfare $\welfare(x,z)$  that is a function of both the production quantity matrix $x$ and the net consumption vector $z$. A  special case of welfare function that is widely used in this framework (see, e.g., \cite[Definition 1]{Johari2005}, or \cite{Cai2019})  is the so-called Walrasian welfare \be\label{Walrasian}\welfare(x,z)=\sum_{j\in\mc M} \int_{0}^{z_{j}} \price_{j}(s) \: \mathrm{d} s-\sum_{i\in\mc N}\cost_{i}\left( x_i\right) \:,\ee that is the difference between the aggregate consumer surplus and the total production cost. 

\begin{remark}\label{remark:multi_markets}
The general form of the production cost functions $\cost_i(x_i)$ that we consider encompasses two scenarios: 
(i) producers incur separate costs for quantities sold in different markets, and (ii) producers face a single, non-separable aggregate cost that depends on all quantities sold across markets (e.g., a weighted sum). The first scenario is particularly relevant for electricity markets, where power can be supplied at a node only by generators physically connected to it, and costs reflect the local plant’s production technology. The second scenario covers the framework considered in network Cournot competition \cite{Abolhassani.ea:2014, Bimpikis2019}, which typically assumes non-separable costs, and more broadly captures decentralized markets in which decentralization arises from fragmented trading venues rather than physical constraints \cite{malamud2017decentralized}. 
For instance,  consider, for a producer $i$ in $\mc N$, the quadratic cost function 
			\be\label{eq:cost-quadratic}\cost_i(x_i)=\sum_{j,k\in\mc M}x_{ij}\costpar_{jk}^{(i)}x_{ik}+\sum_{j\in\mc M}\gamma^{(i)}_{j}x_{ij}\,,\ee
			where $\costpar^{(i)}=(\costpar^{(i)}_{jk})$ in $\R_+^{m\times m}$ is a symmetric nonnegative semi-definite matrix and $\gamma^{(i)}$ in $\R_+^m$ is a nonnegative vector. If $\costpar^{(i)}$ is a nonnegative diagonal matrix, then the cost function of  producer $i$ reduces to $$\cost_i(x_i)=\sum_{j\in\mc M}\left(\costpar^{(i)}_{jj}x_{ij}^2+\gamma^{(i)}_jx_{ij}\right)\,,$$ i.e., the sum of separate quadratic costs of the quantities sold on the different markets. On the other hand, for a rank-$1$ symmetric nonnegative matrix  $\costpar^{(i)}=\eta\gamma^{(i)}(\gamma^{(i)})'$ for some $\eta>0$,  the cost function of  producer $i$ reduces to  $$\cost_i(x_i)=\eta\Big(\sum_{j\in\mc M}\gamma^{(i)}_jx_{ij}\Big)^2+\sum_{j\in\mc M}\gamma^{(i)}_jx_{ij}\,,$$
	i.e., it is a quadratic function of the weighted sum $\sum_{j\in\mc M}\gamma^{(i)}_jx_{ij}$ of the quantities sold on the different markets.
\end{remark}

\subsection{Network Cournot games with  a system operator}

The competition is modeled as a strategic game with $n+1$ players (the $ n $ producers plus the SO) where  every producer $i$ in $\mc N$ chooses a quantity vector $x_i$ in $\mc A_i$ aiming at maximizing its profit, whereas the SO chooses a vector $y$ in $\mc Y$ aiming at maximizing the total welfare $\welfare(x,z)$, where $z$ is as in \eqref{zj}. Precisely, we have the following. 

\begin{definition} A \emph{network Cournot game with a SO (NCGSO)} on a market network \eqref{network}, with production cost functions  \be\label{prod-cost}\cost_i: \R_+^m \to \R_+\,,\qquad i\in\mc N\,, \ee  market inverse demand functions \be\label{price-func}\price_{j}:\R \to \R\,,\qquad j\in\mc M\,,\ee and welfare function \be\label{welfare}\welfare:\R_+^{n\times m}\times\R^{m}\to\R\,,\ee  is a strategic game with player set $\{0\}\cup\mc N=\{0,1,\ldots,n\}$, where the action space of player $0$ (the SO) is $\mc Y$ as defined in \eqref{Y-def},
the action space of every player $i$ in $\mc N$ (the producers) is $\mc A_i$ as defined in \eqref{Ai-def}, and the utility functions are, respectively,  
	\be\label{utility-marketmaker}u_0(y,x)=\welfare(x,\diag(A'x)+By)\,,\ee
	for the SO, and 
	\be\label{utility-producer}\ba{rcl}\!\!u_{i}(x,y)\!\!\!&\!\!\! =\!\!\!&\!\!\ds u_i(x_i,x_{-i},y)\\&\!\!\!=\!\!\!&\!\!\!\!\ds\sum_{j\in\mc M}\!\! A_{i j}x_{ij}  \price_{j}\!\!\left(\sum_{h\in\mc N}\!A_{hj}x_{hj}\!+\!\sum_{k\in\mc L}\!B_{jk} y_k\!\right)\!-\cost_{i}\left(x_{i}\right),\ea\ee
	for every producer $i$ in $\mc N$. 
\end{definition}

\begin{definition} For a NCGSO on a market network \eqref{network}, with production cost functions  \eqref{prod-cost}, market inverse demand functions \eqref{price-func}, and welfare function \eqref{welfare}:
	\begin{enumerate}  \item[(i)] the best response of the SO to the actions $x$ in $\mc X$ of the producers  is $$\mc B_0(x)=\argmax_{ y\in\mc Y}u_0(y,x)\,;$$
		\item[(ii)] the best response of a producer $i$ in $\mc N$ to the actions $x_{-i}$ in $\mc X_{-i}$ of the other producers and the action $y$ in $\mc Y$ of the SO is $$\mc B_i(x_{-i},y)=\argmax_{x_i\in\mc A_i}u_i(x_i,x_{-i},y)\,;$$
		\item[(iii)] a configuration $(x^*,y^*)$ in $\mc X\times\mc Y$ is a (pure-strategy) Nash equilibrium if $$y^*\in\mc B_0(x^*)\,,\qquad x^*_{i}\in\mc B_i(x^*_{-i},y^*)\,,\quad\forall i\in\mc N\,.$$\end{enumerate}\end{definition}
\medskip

\begin{remark}\label{remark:uniqueness} The utility functions of both the SO \eqref{utility-marketmaker} and the producers \eqref{utility-producer} depend on the action $y$ of the SO only through the vector $By$. The market-link incidence matrix typically has $\rank(B)<l$, so that there may exist several actions $y$ in $\mc Y$ with the same image $By$. Clearly, $y\in\mc B_0(x)$ is a best response of the SO to some action $x$ in $\mc X$ of the producers if and only if every $\tilde y$ in $\mc Y$ such that $B\tilde y=By$ is also a best response, i.e.,  $\tilde y\in\mc B_0(x)$.  Symmetrically, $x_i\in\mc B_i(x_{-i},y)$ is a best response of producer $i$ to $(x_{-i},y)$ if and only if  $x_i\in\mc B_i(x_{-i},\tilde y)$ for every $\tilde y$ in $\mc Y$ such that $B\tilde y=By$. It follows that $(x^*,y^*)$ is a Nash equilibrium for a NCGSO if and only if  $(x^*,\tilde y^*)$ is also a Nash equilibrium for every $\tilde y$ in $\mc Y$ such that $B\tilde y=By$. \end{remark}
\medskip
\begin{remark} Maximizing the Walrasian welfare in \eqref{Walrasian}  over $x$ and $z$ under the constraints \eqref{capacity-constraints} and \eqref{z} corresponds to the economic dispatch problem 	that the SO solves in nodal pool markets. In our model, however, production quantities are chosen by producers, and the SO can only influence $z$ through $y$. Therefore, when welfare is given by \eqref{Walrasian}, the SO’s best response is driven solely by the demand side (consumer surplus) and consists of a welfare-maximizing  reallocation of net consumptions $z$, given the injected quantities $x$. This corresponds to the economic dispatch problem in network Cournot models with transmission constraints. Specifically, for affine inverse demand functions $\price_j(z_j)=(d_i-z_j)/a_i$, injected quantities $q_j=\sum_i A_{ij}x_{ij}$ and $H$ as in \eqref{H-electric}, the best response of the SO corresponds to the economic dispatch problem considered in \cite{downward2010cournot} (note that the net consumption in market $i$ is $d_i-x_i$). As proved in \cite{downward2010cournot}, with no capacity constraints, it corresponds to equating the equilibrium prices in every market. \end{remark}
\medskip

\begin{figure}\begin{center}\includegraphics[width=3cm]{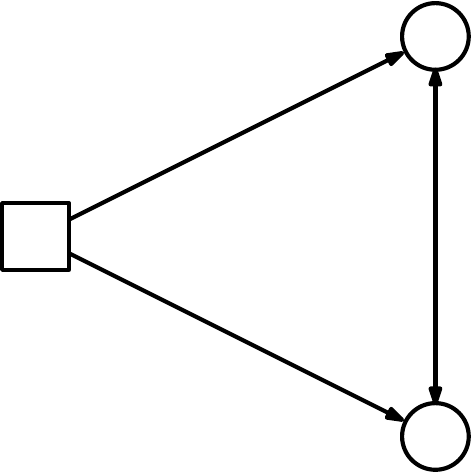}\hspace{1cm}\includegraphics[width=3cm]{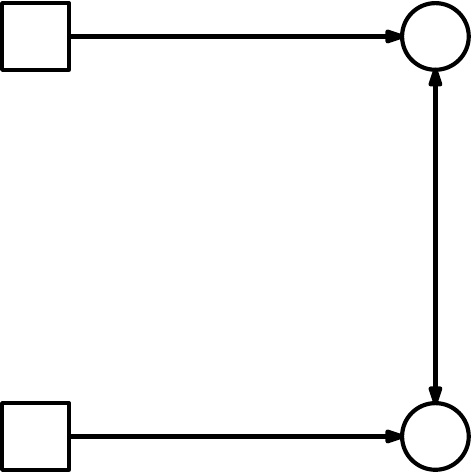}\end{center}\caption{\label{fig:ex-2}On the left: the market network of Example \ref{example:basic}. On the right: an equivalent network for the special case (b) of Example \ref{example:basic}.}\end{figure}

\begin{example}\label{example:basic} Consider a single producer $\mc N=\{1\}$ selling on two distinct markets $\mc M=\{1,2\}$ that are connected by an undirected link with the same capacity $\chi\ge0$ in both directions, so that  $ A=(1,1)$, $B=(-1,1)'$, $c^-=c^+=\chi$. The market network is shown on the left of Figure \ref{fig:ex-2}. Clearly, there are no cycles, so $H=I$. Let both inverse demand functions of the two markets be affine:  \be\label{eq:ex_dem}\price_1(z_1)=1-z_1\,,\qquad\price_2(z_2)=2-2z_2\,.\ee
		Let $z=\diag(A'x)+By=(\tcb{x_{11}}-y,\tcb{x_{12}}+y)\,.$ The SO's utility with \tcb{Walrasian} welfare as in \eqref{Walrasian} is $$\ba{rcl}u_0(y,x)&=&\welfare(x,z)\\&=&z_1+2z_2-z_1^2/2-z_2^2-\cost_1(\tcb{x_{1}})\\&=&\tcb{x_{11}}+2\tcb{x_{12}}-\tcb{x_{11}^2}/2-\tcb{x_{12}^2}-\cost_1(\tcb{x_{1}})\\ &&+(1+\tcb{x_{11}}-2\tcb{x_{12}})y-3y^2/2\,,\ea $$
		and its best response function is \be\label{eq:delta_opt}\mc B_0(x)=\max\left\{-\chi,\min\{\chi,(1+\tcb{x_{11}}-2\tcb{x_{12}})/3\}\right\}\,.\ee
		
		We now consider two special cases, differing in the choice of the production costs.
		
(a) Assume the production cost function to be quadratic as in \eqref{eq:cost-quadratic} with $\costpar^{(1)}=\1\1'$ and $\gamma^{(1)}=0$, so that \be\label{eq:nonsep_costs}\cost_1(\tcb{x_{1}})=(\tcb{x_{11}}+\tcb{x_{12}})^2\,.\ee
		The producer's utility  and best response functions are then $$\ba{rcl}u_1(x,y)\!\!\!&\!\!\!=\!\!\!&\tcb{x_{11}}\price_1(z_1)+\tcb{x_{12}}\price_2(z_2)-\cost(x_{1})\\&\!\!\!=\!\!\!&\tcb{x_{11}}-2\tcb{x_{11}^2}+2\tcb{x_{12}}-3\tcb{x_{12}^2}\\
		&&-2\tcb{x_{11}}\tcb{x_{12}}+(\tcb{x_{11}}-2\tcb{x_{12}})y\,,\ea$$
		and $\mc B_1(y)=\left(\left[\frac{1+5y}{10}\right]_+\wedge\frac{1+y}4,\left[\frac{3-5y}{10}\right]_+\wedge\frac{1-y}3\right)$ respectively. Hence, there is a unique Nash equilibrium $(x^*,y^*)$, $$x^*=\left(1+5y^*,3-5y^*\right)/10\,,\qquad y^*=\min\{\chi,1/3\}\,.$$
Observe that, in equilibrium, there is a positive net flow $y^*$ from market $1$ to market $2$. Indeed, the demand is higher in market $2$ than in market $1$, hence, when the quantity of the good sold on the two markets is the same, the unit price is higher in market $2$ than in market $1$. The best response of the SO aims at reducing the gap between these two prices by moving the net quantity $y^*$ from the market with the lowest price to the one with the highest price. When the link capacity $\chi \geq 1/3$ is sufficiently high, the SO is able to move an optimal quantity $y^*=1/3$ from market $1$ to market $2$, in this way succeeding in equating the unit prices  
		$$\price_1(z^*_1)=\price_1\!\left(\!\frac4{15}\!-\!\frac13\!\right)=\frac{16}{15}=\price_2\!\left(\!\frac2{15}\!+\!\frac13\!\right)=\price_{\tcb{2}}(z^*_2)\,.$$ In contrast, when the link capacity $\chi<1/3$ is not large enough to allow the SO to move the desired quantity of the good, the link from market $1$ to market $2$ saturates and $$\price_1(z^*_1)=\price_1\left((1+5\chi)/10-\chi\right)=9/10+\chi/2\,,$$ $$\price_{\tcb{2}}(z^*_2)=\price_2\left((3-5\chi)/10+\chi\right)=7/5-\chi>\price_1(z^*_1)\,,$$ so that a price difference persists between the two markets for every $0<\chi<1/3$. Finally, it is \tcb{worth} observing that, while $z_2^*=(3+5\chi)/{10}\wedge7/15>0\,,$ for every $\chi>0$, we have that  $z_1^*=(1-5\chi)/10\vee(-1/15)\,,$ that is nonnegative if and only if $\chi\le1/5$. Hence, for every $\chi>1/5$, we have $z_1^*<0$, i.e., the consumption in market $1$ at equilibrium is negative (c.f.~Remark \ref{remark:negative-demand}).
		
(b) Consider now separable quadratic cost functions as in \eqref{eq:cost-quadratic} with $\costpar^{(1)}=I$ and $\gamma^{(1)}=0$, so that \be\label{eq:sep_costs}\cost_1(\tcb{x_{1}})=\tcb{x_{11}^2}+\tcb{x_{12}^2}\,.\ee In this case, the producer's utility and best response are \be\label{utility-sum}u_1(x,y)=\tcb{x_{11}}-2\tcb{x_{11}^2}+2\tcb{x_{12}}-3\tcb{x_{12}^2}+(\tcb{x_{11}}-2\tcb{x_{12}})y\,,\ee and  $\mc B_1(y)=\tcb{\left((1+y)/4,(1-y)/3\right)}$, respectively. Hence, there exists a unique Nash equilibrium $(x^*,y^*)$, where \be\label{Nash-sum}x^*=\left((1+y^*)/4,(1-y^*)/3\right)\,,\qquad  y^*=\chi\wedge7/25\,.\ee 
Therefore, in equilibrium, there is a positive net flow $y^*$ from market $1$ to market $2$. Observe that, when $\chi\ge7/25$, we have \be\label{eq:pr_ex1}\price_1(z^*_1)=\price_1\!\left(\!\frac8{25}\!-\!\frac7{25}\!\right)=\frac{24}{25}=\price_2\!\left(\!\frac6{25}\!+\!\frac7{25}\!\right)=\price_{\tcb{2}}(z^*_2)\,,\ee i.e., the equilibrium prices are the same in the two markets. In contrast, when $\chi<7/25$, so that the link saturates in the direction from market $1$ to market $2$ and $$\price_1(z^*_1)=\price_1\left((1+\chi)/4-\chi\right)=3(1+\chi)/4\,,$$ $$\tcb{\price_2}(z^*_2)=\price_2\left((1-\chi)/{3}+\chi\right)=4(1-\chi)/3>\price_1(z^*_1)\,,$$ so that a price difference persists between the two markets for every $0<\chi<7/25$. In this case, we have that  $z_1^*=(1-3\chi)/4\vee 1/{25}>0$ and $z_2^*=(1+2\chi)/3\wedge{13}/25>0\,,$ so that, at equilibrium, the consumption is positive in both markets for every value of $\chi$. 	
				
Finally, observe that, since in this case the cost \eqref{eq:sep_costs} of the producer is a separable function of the quantities sold in the two markets, the Nash equilibria of this game coincide with the Nash equilibria of a game with two producers with cost functions, respectively, $\cost_1(x_{1})=x^2_{1}$ and $\cost_2(x_{2})=x^2_{2}$, each one selling on a different market (see the network in the right side of Figure \ref{fig:ex-2}). In this latter case, the utilities of the producers are given by \be\label{eq:ut_2p}\begin{aligned} u_1(x_1, x_2, y)&=x_1-2x_1^2+x_1y\\ u_2(x_1, x_2, y)&=2x_2-3x_2^2-2x_2y\,,\end{aligned}\ee (that indeed sum up to the right-hand side of \eqref{utility-sum}) leading to the best responses $\mc B_1(x_2,y)=(1+y)/4$ and $\mc B_2(x_1,y)=(1-y)/3$, and hence to the same Nash equilibrium as in \eqref{Nash-sum}. \end{example}
 
\section{Existence of equilibria and potential}\label{sec3}

In this section, we consider NCGSOs satisfying the following assumption and prove that they always admit a Nash equilibrium. We then consider a subclass of NCGSOs that satisfy additional assumptions and prove that they are exact potential games and that their Nash equilibria are essentially unique. 
\begin{assumption}\label{assumption:cost+price+welfare}\begin{enumerate} \item[(i)] The production cost functions \eqref{prod-cost} are continuous, strictly increasing, convex, and such that $\cost_i(0)=0$; \item[(ii)] the markets' inverse demand functions \eqref{price-func} are continuous, strictly decreasing, concave, and such that \be\label{max-demand}\price_j(\bar z_j)=0\,,\ee  for some finite $\bar z_j>0$;  \item[(iii)] the welfare function \eqref{welfare} is continuous and such that $\welfare(x,z)$ is concave in $z$ in $\R^m$ for every $x$ in $\R_+^{n\times m}$.\end{enumerate}\end{assumption}
Observe that production cost, inverse demand, and welfare functions in Example \ref{example:basic} satisfy Assumption  \ref{assumption:cost+price+welfare}.

\subsection{Existence of Nash equilibria}
We first introduce the following game-theoretic notion. 
\begin{definition} In a NCGSO on a market network \eqref{network}, with production cost functions  \eqref{prod-cost},  market inverse demand functions \eqref{price-func}, and welfare function \eqref{welfare},    an action $x_i$ in $\R_+^{m}$ is \emph{strictly dominated} for a producer $i$ in $\mc N$ if there exists another action $\tilde x_i$ in $\R_+^{m}$ such that $$u_i(x_i,x_{-i},y)<u_i(\tilde x_i,x_{-i},y)\,,$$ for every  action profile $x_{-i}$ in $\R_+^{(n-1)\times m}$ of the other producers and action $y$ in $\mc Y$ of the SO. \end{definition}

\begin{lemma}\label{lemma:strct-dominance} For a producer $i$ in $\mc N$, every $x_i $ in $\R_+^{m}$ such that \be\label{xij<=}A_{ij}=1\,,\qquad x_{ij}>\bar z_j+\sum_{k\in\mc L}|\tcb{B_{jk}}|\max\{c_k^-,c_k^+\}\,,\ee for some market $j$ in $\mc M$ is a strictly dominated action.
\end{lemma}
\begin{proof} Let $j$ in $\mc M$ be a market such that \eqref{xij<=} holds true, and let $\tilde x_i$ in $\mc A_i$ be an action of producer $i$ with entries $\tilde x_{ij}=0$ and $\tilde x_{ig}=x_{ig}$ for every $g$ in $\mc M\setminus\{j\}$. For every $x_{-i}$ in $\mc R_+^{\mc N\setminus\{i\}}$ and $y$ in $\mc Y$, from \eqref{zj}, \eqref{xij<=}, and \eqref{capacity-constraints}, we have that  $$\ba{rclcl}z_j&=&\sum_{h\in\mc N}\!A_{hj}x_{hj}+\sum_{k\in\mc L}\!B_{jk} f_k\\[7pt] &\ge& x_{ij}-\sum_{k\in\mc L}|\tcb{B_{jk}}|\max\{c_k^-,c_k^+\}&>&\bar z_{j}\,,\ea$$ so that   Assumption \ref{assumption:cost+price+welfare}(ii) implies that $\price_{j}\!\left(z_j\right)<\price_{j}(\bar z_j)=0\,.$ It then  follows from  Assumption \ref{assumption:cost+price+welfare}(i) and \eqref{utility-producer} that  $$\ba{rcl}0&<&\ds\cost_i(x_i)-\cost_i(\tilde x_i)-x_{ij}\price_{j}\!\left(z_j\right)\\[7pt]&=&\ds u_i(\tilde x_i,x_{-i},y)-u_i(x_i,x_{-i},y)\,,\ea$$ for every $x_{-i}$ in $\mc R_+^{\mc N\setminus\{i\}}$ and $y$ in $\mc Y$, so that action $\tilde x_i$ strictly dominates action $x_i$. \end{proof}\medskip

\tcb{We also have the following simple result. 
\begin{lemma}\label{lemma:concavity}
Consider a NCGSO on a market network \eqref{network}, with production cost functions  \eqref{prod-cost},  market inverse demand functions \eqref{price-func}, and welfare function \eqref{welfare} satisfying Assumption \ref{assumption:cost+price+welfare}. Then, the utility functions $u_i$ are continuous for $i$ in $\{0\}\cup\mc N$, $r\mapsto \welfare(x,\diag(A'x)+r)$ is concave for $x$ in $\mc X$ and so is $x_i\mapsto u_i(x_i,x_{-i},y)$ for $i$ in $\mc N$, $x_{-i}$ in $\mc X_{-i}$, and $y$ in $\mc Y$. 
\end{lemma}
\begin{proof}We prove only the last statement as the other ones are immediate. 
Let $a_j=\sum_{h\ne i}A_{hj}x_{hj}+\sum_{k}B_{jk} y_k$.
For $x_{ij}$ and $\tilde x_{ij}$ in $\R_+$ and $\lambda$ in $[0,1]$, since $\price_j$ is decreasing, we have 
$$\ba{rcl}
0&\le&\lambda(1-\lambda)(\tilde x_{ij}-x_{ij})(\price_j(x_{ij}+a_j)-\price_j(\tilde x_{ij}+a_j))\\
&=&\bar x_{ij}(\lambda\price_j(x_{ij}+a_j)+(1-\lambda)\price_j(\tilde x_{ij}+a_j))\\
&&-\lambda x_{ij}\price_j(x_{ij}+a_j)-(1-\lambda)\tilde x_{ij}\price_j(\tilde x_{ij}+a_j))\,,
\ea$$
where $\bar x_{ij}=\lambda x_{ij}+(1-\lambda)\tilde x_{ij}$, so that, concavity of $\price_j$ yields
$$\ba{rcl}
&&\bar x_{ij}\price_j(\lambda x_{ij}+(1-\lambda)\tilde x_{ij}+a_j)\\
&\ge&\bar x_{ij}(\lambda \price_j(x_{ij}+a_j)+(1-\lambda)\price_j(\tilde x_{ij}+a_j))\\
&\ge&\lambda x_{ij}\price_j(x_{ij}+a_j)+(1-\lambda)\tilde x_{ij}\price_j(\tilde x_{ij}+a_j))\,,
\ea$$
so that $x_{ij}\mapsto x_{ij}\price_j(x_{ij}+\!a_j\!)$ is concave. Since $\cost_i(x_i)$ is convex, $u_i(x_i,\!x_{-i},\!y)=\sum_{j}\!\! A_{i j}x_{ij}  \price_{j}\!\!\left(x_{ij}\!+\!a_j\!\right)\!-\!\cost_{i}(\!x_{i}\!)$ is the sum of concave functions of $x_i$, hence it is concave in $x_i$. 
\end{proof}}

We are now in a position to prove the first main result of the paper, guaranteeing existence of a Nash equilibrium for NCGSOs satisfying Assumption \ref{assumption:cost+price+welfare}.

\begin{theorem}\label{theorem:Nash-existence} Consider a NCGSO on a market  network \eqref{network}, with production cost functions \eqref{prod-cost}, inverse demand functions \eqref{price-func}, and welfare function \eqref{welfare} satisfying Assumption \ref{assumption:cost+price+welfare}. Then, a Nash equilibrium exists.  \end{theorem}
\begin{proof} Thanks to Lemma \ref{lemma:strct-dominance}, we may  remove the strictly dominated actions and consider the restricted game where the action space of each producer $i$ in $\mc N$ is the hyper-rectangle $$\bar{\mc A}_i=\Big\{v\in\mc A_i:v_j\le A_{ij}\bar z_j+\sum_{k\in\mc L}|\tcb{B_{jk}}|(c_k^-\vee c_k^+)\,,\ \forall j\in\mc M\Big\},$$ that is a non-empty, convex, and compact set.\footnote{\tcb{Indeed, since every action in $\mc A_i\setminus\bar {\mc A}_i$ is strictly dominated by an action in $\bar {\mc A}_i$, no such action can be a best response in the original game. Hence, every Nash equilibrium of the restricted game is also a Nash equilibrium of the original game.}} 

Moreover, following Remark \ref{remark:uniqueness}, we can consider an equivalent game where the action space of the SO is  the set $\mc R=\{r=By:\,y\in\mc Y\}$.  Observe that, thanks to \eqref{BH=B}, we have $\mc R=\{r=Bf:\,f\in\mc F\}$, where $\mc F=\{Hy:y\in\mc Y\}$. Since $\mc F$ is a non-empty, convex, and compact subset of  $\prod_{k\in\mc L}[-c_k^-,c_k^+]$, $\mc R$ is also non-empty, convex, and compact.
Hence, existence of a Nash equilibrium follows from Lemma \ref{lemma:concavity} and a classical result in game theory \cite[Theorem 1.2]{Fudenberg.Tirole:91}.
\end{proof}\medskip

\begin{remark}\label{rem:stack} The existence of Nash equilibria established in Theorem \ref{theorem:Nash-existence} contrasts with results from networked Cournot competition models with transmission constraints, such as \cite{borenstein1997competitive, downward2010cournot}, where the existence of Nash equilibria is in general not guaranteed. The key difference between those models and the one studied here lies in the timing of actions. In our model, producers and the SO act simultaneously, whereas in \cite{borenstein1997competitive, downward2010cournot} the SO first observes the producers’ quantities and then responds optimally. The latter setting is commonly referred to as a Stackelberg game, with producers acting as leaders and the SO as the follower. 
	\tcb{As a result, in our framework, when a producer evaluates its best response, it takes $y$ as given. The markets remain coupled at equilibrium, but a unilateral deviation does not trigger an immediate strategic adjustment by the SO. This implicit coupling via $y$ yields smooth best-response correspondences, since producers do not face instantaneous changes in flows and residual demands induced by their own quantity deviations. In contrast, in \cite{downward2010cournot, borenstein1997competitive}, producers anticipate how their quantities affect the SO’s subsequent decision, and hence network flows and nodal prices. This immediate strategic feedback can increase competitive pressure on producers, at the cost of generating kinks in the residual demand, which may prevent equilibrium existence.}
	A more detailed discussion and comparison of the two frameworks is provided in Section \ref{ss:stack}.\end{remark}

\subsection{Potential game and unique Nash equilibrium}
In this subsection, we determine sufficient conditions for a NCGSO to be an exact potential game. We start with the following definition (c.f.~\cite{MONDERER}).%\cite{Monderer.Shapley:96}). 
\begin{definition}
A NCGSO is an exact potential game if there exists a function $P(x,y)$
to be referred to as the potential function,  such that 
\be\label{dePdexij}P(\tilde x,y)-P(x,y)=u_i(\tilde x,y)-u_i(x,y)\,,\ee
for every $x$ and $\tilde x$ in $\R_+^{n\times m}$ such that $\tilde x_{-i}=x_{-i}$ and every $y$ in $\mc Y$,  
and 
\be\label{dePdeyk}P(x,\tilde y)-P(x,y)=u_0(x,\tilde y)-u_{0}(x,y)\,,\ee
for every $x$ in $\R_+^{n\times m}$ and every $y$ and $\tilde y$ in $\mc Y$. 
\end{definition}

\begin{theorem}
\label{theorem:potential}
Every NCGSO on a market network \eqref{network}, with production cost functions \eqref{prod-cost} satisfying Assumption \ref{assumption:cost+price+welfare}(i), affine inverse demand functions \eqref{price-func} with $\price_j^{\prime}=-\beta_{j}<0$ for every market $j $ in $\mc M$, and Walrasian welfare function \eqref{Walrasian} is an exact potential game with potential function 
\be\label{Pxy}P(x,y)=\welfare(x,\diag(A'x)+By)-\sum_{i\in\mc N} \sum_{j\in\mc M}\frac{A_{ij}\beta_{j}}2x^2_{ij} \,,\ee
and the set of its Nash equilibria coincides with 
\begin{equation} \label{max-potential}
	\argmax_{x\in\mc X,y\in\mc Y} P(x,y)\,.
\end{equation}
Moreover, for every two Nash equilibria $\left(x^{*}, y^{*}\right)$ and $\left(\tilde x^{*},\tilde y^{*}\right)$, we have that $x^*=\tilde x^*$ and $By^*=B\tilde y^*$.
\end{theorem}

\begin{proof} To prove relations \eqref{dePdexij} and \eqref{dePdeyk}, we can equivalently prove that, for every producer $i$ in $\mc N$, the difference $u_i(x, y)-P(x,y)$ is constant in the vector variable $x_i$, and that the difference $u_0(x, y)-P(x,y)$ is constant in the vector variable $y$, i.e., the NCGSO is strategically equivalent to a game with common utility function $P(x,y)$. For every producer $i$ in $\mc N$, we have 
$$\ba{rcl} u_i(x, y)&=&\ds\sum_{j\in\mc M} A_{i j}x_{ij} (\alpha_j-\beta_jz_j)-\cost_{i}\left(x_{i}\right)\\&=&\ds\sum_{j\in\mc M} 
A_{i j}x_{ij} (\alpha_j-\tcb{\beta_jx_{ij}-\beta_j\xi_{ij}})-\cost_{i}\left(x_{i}\right)
%\left(A_{i j}\alpha_jx_{ij} -A_{i j}\beta_jx^2_{ij}\right)-\cost_{i}\left(x_{i}\right) 
%\\&&-\ds\sum_{j\in\mc M} \!\!A_{i j}\beta_jx_{ij}\tcb{\xi_{ij}} %\Big(\sum_{\substack{h\in\mc N\\ h\ne i}}\!A_{hj}x_{hj}+\sum_{k\in\mc L} B_{jk} y_k\Big)
\,,\ea$$ where $\alpha_j=\price_j(0)$ and \tcb{$\xi_{ij}:=\sum_{h\ne i}A_{hj}x_{hj}+\sum_{k\in\mc L} B_{jk} y_k$}, for every market $j$ in $\mc M$. On the other hand,  $P(x,y)$ can be equivalently expressed as
\be\label{potential}\ba{rcl} P(x, y)&=&\ds\sum_{j\in\mc M} \left(\alpha_j \tcb{\left(A_{ij}x_{ij}+\xi_{ij}\right)}-\frac{ \beta_j}{2} \left(A_{ij}x_{ij}+\xi_{ij}\right)^2\right)\\
&&-\ds\sum_{h\in\mc N}\cost_{h}\left(x_{h}\right)-\ds\sum_{h\in\mc N} \sum_{j\in\mc M}\frac{A_{hj}\beta_{j}}2x^2_{hj}\\
&=&\ds\sum_{j\in\mc M}A_{ij}x_{ij}\left( \alpha_j - \tcb{\beta_jx_{ij}-\beta_j\xi_{ij}}\right)-\cost_{i}\left(x_{i}\right)\\
&&+\ds\sum_{j\in\mc M}\left(\alpha_j\xi_{ij}-\frac{ \beta_j}{2}\xi_{ij}^2\right)-\tcb{\zeta_{i}}%\\
%&&-\ds\sum_{j\in\mc M} A_{i j}\beta_jx_{ij} \Bigg(\sum_{\substack{h\in\mc N\\ h\ne i}}\!A_{hj}x_{hj}+\sum_{k\in\mc L} B_{jk} y_k\Bigg)\\[-7pt]
%&& {\color{red}+\ds\sum_{j\in\mc M} \alpha_j \Bigg(\sum_{\substack{h\in\mc N\\ h\ne i}}\!A_{hj}x_{hj}+\sum_{k\in\mc L} B_{jk} y_k\Bigg)}\\
%&&-\ds\sum_{j\in\mc M}\frac{ \beta_j}{2} \Bigg(\sum_{\substack{h\in\mc N\\ h\ne i}}\!A_{hj}x_{hj}+\sum_{k\in\mc L} B_{jk} y_k\Bigg)^2\\
%&&
\,,\ea\ee
\tcb{where $\zeta_i:=\sum_{ h\ne i}(\cost_{h}\left(x_{h}\right)- \sum_{j\in\mc M}\frac{A_{hj}\beta_{j}}2x^2_{hj})$.
Since neither $\xi_{ij}$ nor $\zeta_i$ depends on $x_i$, also} the difference 
\be\label{ui-P}
u_i(x,y)-P(x,y)=\sum_{j\in\mc M}\left(\alpha_j\xi_{ij}-\frac{ \beta_j}{2}\xi_{ij}^2\right)-\ds\zeta_{i}\,,
\ee
does not depend on $x_i$. On the other hand, it directly follows from \eqref{utility-marketmaker} and \eqref{Pxy} that 
\be\label{u0-P}u_0(x,y)-P(x,y)=\sum_{i\in\mc N} \sum_{j\in\mc M}\frac{A_{ij}\beta_{j}}2x^2_{ij}\,,\ee is constant in $y$. 
\tcb{ This proves that the NCGSO is an exact potential game with potential function $P(x,y)$ and thus that
the Nash equilibria of the NCGSO coincide with the Nash equilibria of a game with common utility function $P(x,y)$.} 

\tcb{As a consequence, global maximum points of the potential function $P$ are Nash equilibria. We are then left with proving that Nash equilibria are necessarily maximum points of $P$. Observe that the potential function $P$ as defined in \eqref{potential} is a concave function.  When all production cost functions $\cost_i$ are $\mc C^1$, so is $P$, and the result is standard (e.g., see \cite{neyman1997concave} at page 226). Below, we show that the result remains true when the production cost functions $\cost_i$ are not necessarily $\mc C^1$. }

{\color{black}We use the concept and basic properties of sub-gradients \cite{rock1970convex}. Recall that, for a nonempty convex set $D\subseteq\R^k$ and a convex function $f:D\subseteq \R^k\to\R$, a vector $h$ in $\R^k$ is a sub-gradient of $f$ at  $z^*$ in $D$ if $f(z)\geq f(z^*)+h'(z-z^*)$ for every $z$ in $D$. The set of the sub-gradients of $f$ in $z^*$ is denoted by $\partial f(z^*)$. %If $f$ is $\mc C^1$, then the differential of $f$ at $z^*$ is its only sub-gradient.
%$\partial f(z^*)$ is a singleton consisting of the differential at $z^*$. 
Notice that $0\in \partial f(z^*)$ if and only if  $z^*$ is a global minimum point of $f$.
When variables in the domain split, say $z=(x,y)$, we indicate with $\partial_{x}f(x^*,y^*)$, the set of sub-gradients of $f$ in $(x^*,y^*)$, when $f$ is thought as a function exclusively of $x$.
}
%
%
%$f(x,y)$, we indicate with $\partial f(x^*,y^*)\subseteq\R^{nm+l}$ the set of its sub-gradients \cite{rock1970convex} in the point $(x^*, y^*)$. Furthermore, we indicate with $\partial_{x_i} f(x^*,y^*)\subseteq\R^{m}$ and $\partial_{y}f(x^*,y^*)\subseteq\R^{l}$, the sub-gradients of $f$ in $(x^*,y^*)$, when $f$ is thought as a function exclusively of only $x_i$ and, respectively, of only $y$. 
\tcb{From expression \eqref{potential}, we represent $$-P(x,y)=Q(x, y)+\Psi(x)\,,$$ where $Q$ is $\mc C^1$ and convex, and $\Psi(x)=\sum_{i\in\mc N}\Psi_i(x_i)$ where each function $\Psi_i$ is convex. }
{\color{black}If $(x^*,y^*)$ is a Nash equilibrium, then the functions  $x_i\mapsto -P(x_i,x^*_{-i}, y^*)$ for $i$ in $\mc N$, and $y\mapsto -P(x^*, y)$ have a minimum point in $x^*_i$ and $y$, respectively. Consequently,  $0\in\partial_{x_i}(-P)(x^*, y^*)$ for every $i$ in $\mc N$ and $0\in\partial_{y}(-P)(x^*, y^*)$. 
Additivity properties of sub-gradients \cite[Theorem 23.8]{rock1970convex}   yield $\partial_{y}(-P)(x^*, y^*)=\partial_{y}(Q)(x^*, y^*)$ and 
$\partial_{x_i}(-P)(x^*, y^*)=\partial_{x_i}(Q)(x^*, y^*)+\partial_{x_i}(\Psi_i)(x^*_i)$  for every $i$ in $\mc N$. 
This implies that $0\in\partial_{y}Q(x^*, y^*)$ and, for every $i$ in $\mc N$, there exists a vector $v_i$ such that  
$v_i\in\partial_{x_i}Q(x^*, y^*)$ and $-v_i\in\partial_{x_i}(\Psi_i)(x^*_i)\,.$ 
Now, let $v=(v_1,v_2,\ldots,v_n)$. {\color{black}The fact that $Q$ is $\mc C^1$, together with the characterization of sub-gradients in \cite[Theorem 23.2]{rock1970convex}, yields $\nabla_{x_i}Q(x^*, y^*)' z_i\geq v_i' z_i$ for every vector $z_i$ in $\R^m$ pointing inside $\mc A_i$ (i.e., such that $x^*_i+\eps z_i\in\mc A_i$ for sufficiently small $\eps$) and similarly, $\nabla_{y}Q(x^*, y^*)'w\geq 0$ for every vector $w$ in $\R^l$ pointing inside $\mc Y$. Summing up, we obtain that $\nabla Q(x^*, y^*)' (z, w)\geq (v,0)' (z,w)$ for every $(z,w)$ pointing inside $\mc X\times\mc Y$. Using again \cite[Theorem 23.2]{rock1970convex}, we deduce that $(v,0)\in\partial_{}Q(x^*, y^*)$. On the other hand, the separable structure of $\Psi$ yields
$(-v, 0)\in\partial_{}\Psi(x^*)$.}
%
%
%\tcr{Then, the fact that $Q$ is $\mc C^1$ and the separable structure of $\Psi$ yield 
%$$(v,0)\in\partial_{}Q(x^*, y^*)\,,\qquad  (-v, 0)\in\partial_{}\Psi(x^*)\,.$$}
Using again the additivity property \cite[Theorem 23.8]{rock1970convex}, we finally obtain that $0=(v, 0)+(-v, 0)\in\partial(-P)(x^*, y^*)\,,$ so that $(x^*, y^*)$ is a minimum point of $-P$ and thus a maximum point of $P$.}

Finally, the last point  of the claim follows from the fact that $P$ is actually strictly concave in the variables $(x, By)$\tcb{, since the quadratic term $-\sum_{ij}A_{ij}\beta_jx_{ij}^2/2$ is strictly concave in $x$, while the aggregate consumer surplus term $\sum_j\int_0^{z_j}\price(s)\de s$ is strictly concave in $z=\diag(A'x)+By$.}
\end{proof}\medskip

\begin{example}
	Consider the same setting as in Example \ref{example:basic} (a), that is, one producer with the quadratic cost in \eqref{eq:nonsep_costs} that sells on two markets with inverse demand functions as in  \eqref{eq:ex_dem}. If we consider the \tcb{Walrasian} welfare, then the assumptions of Theorem \ref{theorem:potential} are satisfied. Starting from the computations in Example \ref{example:basic}, we obtain that the potential function in \eqref{Pxy} is 
	$$
	\begin{aligned}
		P(x,y)=	&u_0(y,x)- \frac{1}2x^2_{11}-x^2_{12}\\=&x_{11}+2x_{12}-2x_{11}^2-3x_{12}^2-2x_{11}x_{12}\\ &+(1+x_{11}-2x_{12})y-3y^2/2\,.
	\end{aligned}
	$$
%	Observe that there holds
%	$$
%	\ds\frac{\partial P}{\partial x_{11}}(x,y)=\ds\frac{\partial u_1}{\partial x_{11}}(x,y)=1+4x_{11}-2x_{2}+(y_1-y_2)
%	$$
%	and
%	$$
%	\ds\frac{\partial P}{\partial x_{2}}(x,y)=\ds\frac{\partial u_1}{\partial x_{2}}(x,y)=2+6x_{11}-2x_{2}-2(y_1-y_2)
%	$$
%	Also, we have that
%	$$
%	\ds\frac{\partial P}{\partial y_{k}}(x,y)
%	=\frac{\partial u_0}{\partial y_{k}}\left(y,x\right)\,,
%	$$
%	The Nash equilibrium obtained in Example \ref{example:basic} can be found by computing \eqref{max-potential}.
\end{example}

\section{Price differences, link saturation, and non-negativity of the demand} \label{sec4}

In this section, we investigate the relationship between price differences and link saturations of the physical flow $f=Hy$ induced by the SO when playing a best response \be\label{current-BR}y\in\mc B_0(x)\,,\ee and the capacity constraints \eqref{capacity-constraints}. Finally, we explore their consequences on the sign of the consumptions in the markets. 

%Before proceeding, it is worth pointing out that our results hold true for NCGSOs in every configuration $(x,y)$ such that the action of the SO $y$ in $\mc B_0(x)$ is a best response to those of the producers. Hence, in particular, the results in this section apply to every Nash equilibrium, as well as in more general configuration whereby some producers might not be playing best response.  

\subsection{System operator's best response and link saturation}
In order to ease readability, we formulate the results in this subsection for the special case of Walrasian welfare\footnote{However, it is evident that all the results in this subsection hold true ---with the very same proofs--- for every welfare function $\welfare(x,z)$ that is differentiable in the consumption vector $z$, provided that one replaces the prices $\price_j(z_j)$ in each market $j$ in $\mc M$ with the partial derivative $\frac{\partial\welfare}{\partial z_{j}}(x,z)$ of the welfare function with respect to the consumption $z_j$ in the same market.} function \eqref{Walrasian}.  We first introduce the vector  $\Delta=B'\price(z)$  whose entries \be\label{Deltak}\Delta_k=\price_{\tau_k}(z)-\price_{\sigma_k}(z)\,,\ee correspond to \tcb{the} difference between the prices at the end nodes of each link $k$ in $\mc L$. The following general result holds.  

\begin{theorem}\label{lemma2} Consider a NCGSO on a market network \eqref{network}, with production cost functions \eqref{prod-cost}, inverse demand functions \eqref{price-func}, and Walrasian welfare \eqref{Walrasian}. For $x$ in $\mc X$ and $y$ satisfying \tcb{\eqref{current-BR}}, let $f=Hy$, $z=\diag(A'x)+By$, and $\Delta=B'\price(z)$. Then, for every link $k$ in $\mc L$ such that $\Delta_k\ne0$, there exists a link $j$ in $\mc L$ such that either 	 \be\label{eq:pricederiv-1} H_{jk}\Delta_k>0 \,,\qquad	f_j=c^+_j\,,\ee or  \be\label{eq:pricederiv-2}H_{jk}\Delta_k<0 \,,\qquad	f_j=-c^-_j\,.\ee \end{theorem}
\smallskip
\begin{proof}
Consider a link $k$ in $\mc L$ such that $\Delta_k>0$  and assume by contradiction that $(Hy)_j<c_j^+$ for every link $j$ in $\mc L$ such that $H_{jk}>0$ and $(Hy)_j>-c_j^-$ for every link $j$ in $\mc L$ such that $H_{jk}<0$. 
Then, there would exist $\overbar{\eps}>0$ such that \be\label{Hy+eps<c}-c^-\le H(y+\eps\delta^{(k)})=Hy+\eps H_{\cdot k}\le c^+\,,\ee for every $\eps$ such that  $0 \le \eps \le \overbar{\eps}$, and
\be\label{de>0}\hspace{-0.3cm}\ba{rcl}\ds\left.\frac{\de}{\de \eps} u_0\left(y+\eps\delta^{(k)},x\right)\right|_{\eps=0} 
\!\!\!\!\!\!& = &\!\!\!\!\!\!\ds\left.\frac{\de}{\de \eps} \welfare\left(x,z+\eps B\delta^{(k)}\right) \right|_{\eps=0}\\
			&\!\!\!=\!\!\!&\!\!\!\!\!\!\ds	\left. \frac{\de}{\de \eps} \welfare\left(x,z+\eps \delta^{(\tau_k)}-\eps \delta^{(\sigma_k)}\right)\right|_{\eps=0}\\&\!\!\!=\!\!\!&\!\!\!\!\!\!\ds\frac{\partial\welfare}{\partial z_{\tau_k}} \left(x,z\right)-\frac{\partial\welfare}{\partial z_{\sigma_k}}\left(x,z\right)\\&\!\!\!=\!\!\!&\!\!\!\price_{\tau_k}(z)-\price_{\sigma_k}(z)\\&\!\!\!=\!\!\!&\!\!\!\Delta_k\\&\!\!\!>\!\!\!&\!\!\!0\,,\ea\ee
where the first equality follows from \eqref{utility-marketmaker}, the second one from \eqref{Bij}, the fourth one from the definition \eqref{Walrasian} of the Walrasian welfare function, and the fifth one from \eqref{Deltak}.
Inequalities \eqref{Hy+eps<c} and \eqref{de>0} together imply that we can find $\eps^*>0$ such that $$y+\eps^*\delta^{(k)}\in\mc Y\,,\qquad u_0(y,x) < u_0(y+\eps^*\delta^{(k)},x)\,,$$ thus contradicting assumption \eqref{current-BR}. This proves that there must exist a link $j$ in $\mc L$ such that either \eqref{eq:pricederiv-1} or \eqref{eq:pricederiv-2} hold true. 

%\tcb{The case $\Delta_k<0$ follows analogously by considering the perturbation $y-\eps\delta^{(k)}$.}
% .
%
Arguing similarly, we find that, for a link $k$ in $\mc L$ such that $\Delta_k<0$, 
if by contradiction $(Hy)_j<c^+_j$ for every link $j$ in $\mc L$ such that $H_{jk}<0$ and $(Hy)_j>-c_j^-$ for every link $j$ in $\mc L$ such that $H_{jk}>0$, then  there would exist $\overbar{\eps}>0$ such that $-c^-\le H(y-\eps\delta^{(k)}) \le c^+\,,$ for every $\eps$ such that  $0 \le \eps \le \overbar{\eps}$, and
$\left.\frac{\de u_0}{\de \eps} \left(y-\eps\delta^{(k)},x\right)\right|_{\eps=0}=-\Delta_k>0\,,$
which implies that we can find $\eps^*>0$ such that $y-\eps^*\delta^{(k)}\in\mc Y$ and  $u_0(y,x) < u_0(y-\eps^*\delta^{(k)},x)\,,$ thus contradicting assumption \eqref{current-BR}.  This shows that also in this case there must exist a link $j$ in $\mc L$ such that either \eqref{eq:pricederiv-1} or \eqref{eq:pricederiv-2} hold true, thus completing the proof.
\end{proof}\medskip
Theorem \ref{lemma2} states that, when the SO is playing a best response, if the  prices differ across the endpoints of a link $k$, then there exists a link $j$ such that $H_{jk}\ne0$ and the physical flow $f_j$ is saturated, i.e., one capacity constraint binds as in \eqref{eq:pricederiv-1} or \eqref{eq:pricederiv-2}. 

\begin{figure}
	\centering
	
	\includegraphics[width=0.26\textwidth]{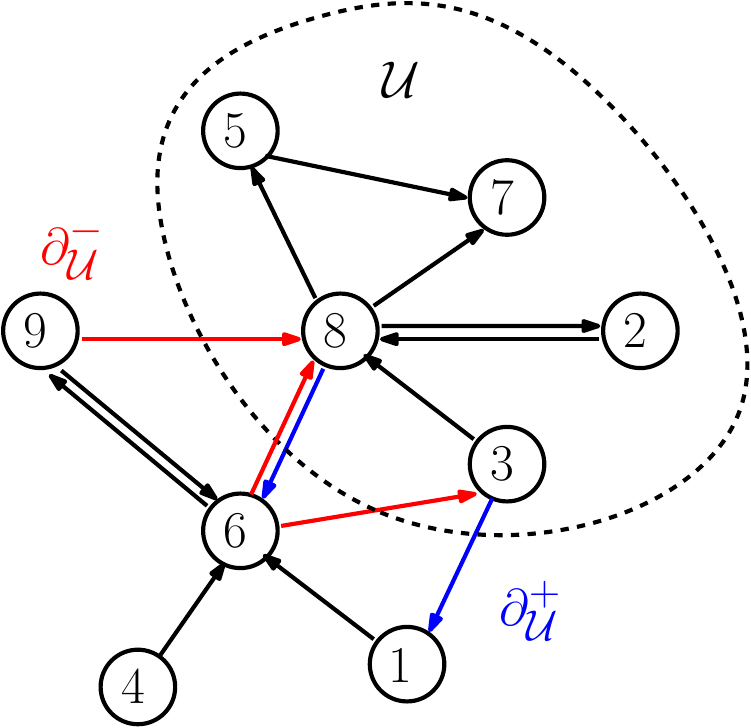}
	\caption{A cut in a network.} %\tcr{FIGURA DA CAMBIRARE O PER LO MENO AGGIORNARE}}
\label{fig2}
\end{figure}

Clearly, in the special case when $H=I$ is the identity matrix, the only link $j$ such that $H_{jk}\ne0$ is link $k$ itself (with $H_{kk}=1$), so that Theorem \ref{lemma2} implies that every link $k$ across which there is a price difference is saturated in the direction from the market with the lower price to the market with larger price. This is consistent, e.g., with what was observed at the Nash equilibrium in Example \ref{example:basic}.
In fact in the special case when $H=I$,  Theorem \ref{lemma2} implies the existence of critical cuts with mono-directional saturated flow, as stated below.  For a subset of markets $\mc U\subseteq\mc M$ define the in- and out-boundaries (c.f.~Figure \ref{fig2})
\tcb{$$ \partial_{\mc U}^-   = \left\{ k \in \mc L \: : \: \sigma_k\in\mc M\setminus\mc U\,,\   \tau_k\in \mc U      \right\}\,,$$
$$\partial_{\mc U}^+   = \left\{ k \in \mc L \: : \:  \sigma_k \in \mc U\,,\ \tau_k\in \mc M\setminus\mc U  \right\}\,.$$}
We then have the following.
\begin{corollary}
	\label{cor:pircemiss}
Consider a NCGSO on a market network \eqref{network} with $H=I$, production cost functions \eqref{prod-cost}, inverse demand functions \eqref{price-func}, and Walrasian welfare \eqref{Walrasian}.   
	For $x$ in $\mc X$ and $y$ satisfying \tcb{\eqref{current-BR}}, let $z=\diag(A'x)+By$. Then, for every nonempty proper subset of markets $\mc U\subseteq\mc M$ such that \be\label{eq:pricemissc}\max_{j \in \mc U} \price_j(z_j)  <  \min_{j \in \mc M\setminus\mc U} \price_j(z_j)\,,\ee we have \be\label{eq:sat} y_{k}=\left\{\ba{lcl}\tcb{-c^{-}_k}&\se& k\in\tcb{\partial^-_{\mc U}}\,,\\[3pt]\tcb{c_k^+}&\se& k\in\tcb{\partial^+_{\mc U}}\,.\\\ea\right.\ee
\end{corollary}\smallskip
\begin{proof} Equation \eqref{eq:pricemissc} implies that $\Delta_k>0$ for every link $k$ in the out-boundary $\partial_{\mc U}^+$ and  $\Delta_k<0$ for every link $k$ in the in-boundary $\partial_{\mc U}^-$. Equation \eqref{eq:sat} then follows from \eqref{eq:pricederiv-1} and \eqref{eq:pricederiv-2}, since $H=I$. 
\end{proof} \medskip

Recall that, in the special case when the market network contains no cycles, we have that necessarily $H=I$, so that in this case Corollary \ref{cor:pircemiss} always applies, implying that if $\price_j(z_j)<\price_{j^*}(z_{j^*})$ for two markets $j$ and $j^*$ in $\mc M$, then there exists a link $k$ along the unique path connecting $j$ and $j^*$ such that  either $f_k=c^+_k$ (if link $k$ is oriented in the direction followed by the path from $j$ to $j^*$) or $f_k=-c^-_k$ (if link $k$ is oriented in the opposite direction). 

In contrast, when the market network contains cycles, as already noted, there are several matrices $H$ satisfying the identity \eqref{BH=B} and which links saturate depends on the matrix $H$, as the following example illustrates. 

\addtocounter{example}{-3}
\begin{example}[cont'd] 
Consider the market network in Figure \ref{fig:ex-1}. Let the inverse demand functions be affine %as in %\eqref{eq:price-affine}
	\be\label{eq:price-affine}\price_j(z_j)=\alpha_j-\beta_jz_j\,,\ee
		 with $\alpha_j=10$ for $j=1,2,3$, $\alpha_4=5$, and $\beta_j=1$ for every market $j$ in $\mc M$. %\tcr{(The compact notation might be new. Otherwise  $\beta_j=1$ for all $j$ and $\alpha_j=10$ for all $j\leq 3$ and $\alpha_4=5$,}
Let $x_1=(1,0,0,0,0)$, $x_2=(0,9,0,0)$, and $x_3=(0,0,5,0)$. 
%$$x=\left(\ba{ccccc}1&0&0&0\\0&9&0&0\\0&0&5&0 \ea\right).$
When $H=I$, the best response is $$\mc B_0(x)=\{(2,-2,1-\alpha,1+\alpha,2):0\le\alpha\le1\}\,,$$  corresponding to the SO sending $4$ units of flow from market $2$ to market $1$, $2$ of which through the direct link $2$, and the other two through the path $(2,3,4,1)$, in this way achieving uniform prices $\price_j(z_j)=5$ for every market $j$ in $\mc M$. In contrast, when $H$ is given by \eqref{H=4}, then the SO can send at most $11/3$ units of flow from market $2$ to market $1$. This is achieved through any best response 
$$
y\in \mc B_0(x)=\left\{y\in \R^{l}\,\mid \, Hy=(2,-5/3,1,1,2)\right\}\,,
$$
and leads to a consumption vector $z=({14}/{3},{16}/{3},5,0)$ and non-uniform prices $\price_1(z_1)={16}/{3}$, $\price_2(z_2)={14}/{3}$, and $\price_3(z_3)=\price_4(z_4)=5$. In this case, there are three saturated links $k=1,3,5$. Link $k=2$ is not saturated, but its physical flow $f_2$ cannot be increased without increasing the flows on the saturated links that lie on the indirect path $(2,3,4,1)$ from market $2$ to market $1$. 
\end{example}
\addtocounter{example}{+5}

\subsection{Non-negativity of the demand} \label{sec:positive}
As mentioned in Remark \ref{remark:negative-demand}, the demand vector $z$ can have negative entries in general. Enforcing the constraint $z \ge 0$ while computing the SO's best response comes at the cost of complicating the model, requiring the analysis of the resulting generalized Nash equilibria of the game. The following result provides a sufficient condition for the non-negativity of $z$.

	\begin{theorem}
		\label{theo:pos}
		Consider a NCGSO on a market network \eqref{network} \tcb{with $H=I$}, with production cost functions \eqref{prod-cost}, inverse demand functions \eqref{price-func} \tcb{and Walrasian welfare function \eqref{Walrasian}} satisfying Assumption \ref{assumption:cost+price+welfare}. 
If  \begin{equation}\label{eq:ass_pos}
\price_j(0) = a\,,\qquad  \forall \: j \in \mc M\,,
\end{equation} 
		for some constant $a>0$, then 
		\be\label{z>=0}z=\diag(A'x)+By\ge0\,,\ee 
		for every $x$ in $\R_+^{n\times m}$ and $y$ in $\mc B_0(x)$.  	
	\end{theorem}
	\begin{proof}
Consider the subset of markets $\mc U = \{j \in \mc M \: : \: z_j \ge 0 \}$. 
If $\mc U$ were empty, then we would have 
$$
0>\sum_{j\in\mc M}z_j=\1'z=\1'\diag(A'x)+\1'By=\sum_{i\in\mc N}\sum_{j\in\mc M}x_{ij}\ge0\,,
$$
which is a contradiction. Hence, necessarily $\mc U$ is nonempty. Observe that, by Assumption \ref{assumption:cost+price+welfare}(ii), the inverse demand functions are all strictly decreasing so  that 
\be\label{<a}\price_{j}(z_{j})>\price_j(0)= a\,,\qquad \forall j\in\mc M\setminus\mc U\,,\ee 
\be\label{>a}\price_{j}(z_{j})\le\price_j(0)= a\,,\qquad \forall j\in\mc U\,.\ee 

Assume by contradiction that  $\mc U\ne\mc M$. Let	$v\in\mc M\setminus\mc U$. Then, \eqref{<a} and \eqref{>a} would imply  \eqref{eq:pricemissc}, so that Corollary \ref{cor:pircemiss} would imply that  \eqref{eq:sat} holds true. In turn, this would yield $$\sum_{j\in\mc M\setminus\mc U}(By)_{j}  = \tcb{-} \sum_{k\in \partial_{\mc U}^{\tcb -} }y_k\tcb{+}\sum_{k\in\partial_{\mc U}^{\tcb+}}y_k =  \sum_{k\in \partial_{\mc U}^{\tcb{-}}}c_k^{\tcb{-}}+\sum_{k\in \partial_{\mc U}^+}c_k^{\tcb{+}}\ge0\,,$$ so that \eqref{z} implies that $$0>\sum_{j\in\mc M\setminus\mc U}z_j=\sum_{j\in\mc M\setminus\mc U}(By)_{j}+\sum_{j\in\mc M\setminus\mc U}\sum_{i\in\mc N}A_{ij}x_{ij}\ge0\,,$$which is a contradiction. Therefore, it must be $\mc U=\mc M$, thus proving the claim.
	\end{proof}

	\medskip
		
		\begin{example}\label{example:basic2}
			{Let us now consider the same setting as in Example \ref{example:basic}, except for the inverse demand functions of the two markets that we now set to be affine as in \eqref{eq:price-affine} with $\alpha_1=\alpha_2=1$, $\beta_1=1$, and $\beta_2=2$, so that  
			$\price_1(z_1)=1-z_1$ and $\price_2(z_2)=1-2z_2\,.$
			Observe that $\price_1(0)=\price_2(0)=1$ and therefore condition \eqref{eq:ass_pos} is satisfied.  Let $z=\diag(A'x)+By=(x_{11}-y,\tcb{x_{12}}+y)\,.$ The SO's utility is then given by
			$$
			\ba{rcl}u_0(y,x)&=&\welfare(x,z)\\		&=&z_1+z_2-z_1^2/2-z_2^2-(x_{11}+\tcb{x_{12}})^2\\
			&=&x_{11}+\tcb{x_{12}}-3x_{11}^2/2-2x_{12}^2-2x_{11}x_{12}\\
			&&+(x_{11}-2x_{12})y-3y^2/2\,,\ea
			$$
			and its best response function is
$$\mc B_0(x)=(-\chi)\vee\left((x_{11}-2x_{12})/3\wedge\chi\right)\,.$$
			Observe that, for every $x$ in $\mc X$, 
			$$
z_1^*=x_{11}-\mc B_0(x)%\Delta^*(x)%=x_{11}+\frac{2x_{12}-x_{11}}{3}
=\begin{cases}
	x_{11}-\chi &\hspace{-0.3cm}\se\ds x_{11}-2x_{12}\ge3\chi\\ \frac23(x_{11}+x_{12})&\hspace{-0.3cm}\se\ds{|x_{11}-2x_{12}|}<3\chi\\
x_{11}+\chi&\hspace{-0.3cm}\se\ds x_{11}-2x_{12}\le-3\chi\,,\end{cases}$$

$$z^*_2=x_{12}+\mc B_0(x)%\tcr{\Delta^*(x)}%=x_{12}-\frac{2x_{12}-x_{11}}{3}
=\begin{cases}
	x_{12}+\chi&\hspace{-0.3cm}\se\ds x_{11}-2x_{12}\ge3\chi\\ \frac13(x_{11}+x_{12})&\hspace{-0.3cm}\se\ds{|x_{11}-2x_{12}|}<3\chi\\
x_{12}-\chi&\hspace{-0.3cm}\se\ds x_{11}-2x_{12}\le-3\chi\,,
	\end{cases}$$
			so that both $z_1^*\ge0$ and $z_2^*\ge0$, as predicted by Theorem \ref{theo:pos}.}
	\end{example}	\medskip
	
	\begin{remark}In contrast to Example \ref{example:basic2}, in Example \ref{example:basic} we had $\price_1(0)=1\neq \price_2(0)=2$ and therefore \eqref{eq:ass_pos} was not satisfied. Notice that the equilibrium demand $z^*_1$ can be negative in that case.  Observe that Theorem \ref{theo:pos} gives a sufficient condition, and indeed there are some cases where $z^*$ is nonnegative even if the condition is not satisfied (see Example \ref{example:basic} (b), for instance). \end{remark}

{\color{black}
		Notice that the conclusion of Theorem~\ref{theo:pos} may not hold in the general case when $H\neq I$, as shown in the following example.

\begin{example}\label{example:negativeH}
Consider the market network $(A,B,c^-,c^+,H)$ of Example~\ref{example:1}, 
with the matrix $H$ as in \eqref{H=4}. Let the inverse demand functions be affine as in \eqref{eq:price-affine} with
$\beta_1=6$, $\beta_2=1,$  $\beta_3=\beta_4=1/5$, and $\alpha_j=12$ for every market $j$ in $\mc M$, and the welfare function be the Walrasian \eqref{Walrasian}.  
%In particular, condition \eqref{eq:ass_pos} is satisfied with  $a=12$. 
Consider the production profiles
$x_1^*=(2/5,0,0,0)\,,$ $x_2^*=(2/5,3/2,0,0)\,,$
$x_3^*=(0,3/2,1,0)\,$  so that 
$q^*:=\diag(A'x^*)=(4/5,3,1,0)\,,$
and the flow
$f^*=\frac1{513}(0,-13,368,368,-513)$,
which satisfies both Kirchhoff's second law \eqref{eq:rangeH} and 
the capacity constraints \eqref{capacity-constraints}. %: the bounds  $f_1\ge-c^-_1=0$ and $f_5\ge-c^-_5=-1$ are active, while all the  remaining ones are slack. 
Indeed, 
$y^*=\frac1{513}(0,-13,736,0,-513)$ is such that $f^*=Hy^*$, so that $y^*\in\mc Y$ is a feasible action of the SO.  
The corresponding consumption profile is
$z^*=q^*+Bf^*=(\frac{2117}{2565}, \frac{790}{513},
\frac{1762}{513}, -1)\,,$
and the market prices are
$\price_1(z^*_1)=\frac{6026}{855}$,  
$\price_2(z^*_2)=\frac{5366}{513}$, 
$\price_3(z^*_3)=\frac{29018}{2565}$, and 
$\price_4(z^*_4)=\frac{61}{5}$, respectively, 
so that the price-difference vector \eqref{Deltak} is
$\Delta^*=B'\price(z^*)
=\frac1{2565}(-13215,8752,2188,2188,2275)$.
Consider linear production cost functions 
$\cost_i(x_i)=\sum_{j\in\mc M}\rho_j\,x_{ij}$, %  for every producer $i$ in $\mc N$, 
with\footnote{Notice that, for every two producers $h$ and $i$ in $\mc N$ and market $j$ in $\mc M$, we have $x^*_{hj}=x^*_{ij}$ 
and $\rho_j>0$, so that Assumption~\ref{assumption:cost+price+welfare}(i) is satisfied.} 
\be\label{FOC-x}\rho_j=\price_j(z^*_j)-\beta_j\,x^*_{ij}\,,\qquad \forall j\in\mc M,i\in\mc N\,.\ee
Then, Theorem~\ref{theorem:potential} implies that the Nash 
equilibria of this NCGSO coincide with the maximum points over 
$\mc X\times\mc Y$ of the potential function \eqref{Pxy} which 
is a concave quadratic function. 
%Since $\mc X\times\mc Y$ is convex, in order to prove that 
%$(x^*,y^*)$ is a maximum point of $P$ it is sufficient to verify the 
%first-order condition
%$\nabla P(x^*,y^*)'((x,y)-(x^*,y^*))\le0$ for every $x$ in $\mc X$ and $y$ in  $\mc Y$. 
On the one hand, Equation \eqref{FOC-x} implies that 
$\frac{\partial P}{\partial x_{ij}}(x^*,y^*)
=\price_j(z^*_j)-\beta_j\,x^*_{ij}-\rho_j=0$ if $A_{ij}=1$, so that $\nabla P(x^*,y^*)'(x-x^*)=0$ for every $x$ in $\mc X$. On the other hand, for $y$ in $\mc Y$, 
%we have that 
$$\ba{rcl}
\nabla_yP(x^*,y^*)'(y-y^*)
&=&\price(z^*)'B(y-y^*)\\
&=&\price(z^*)'BH(y-y^*)\\
&=&(\Delta^*)'(f-f^*)\\ 
&=&(\Delta^*_1-{\Delta^*_2}/4)(f_1-f_1^*)\\
&&+(2\Delta^*_3-{\Delta^*_2}/2)(f_3-f_3^*)\\
&&+(\Delta^*_5-{\Delta^*_2}/3)(f_5-f_5^*)\\
&\le&0\,,
\ea$$
where the first identity follows from \eqref{Walrasian} and \eqref{z}, the second one from \eqref{BH=B}, the third one from \eqref{Deltak} and $f=Hy$, and the fourth one from \eqref{eq:rangeH}, while the final inequality holds since $\Delta^*_1\le{\Delta^*_2}/4$, $f_1\ge -c_1^-=f_1^*$, $2\Delta^*_3={\Delta^*_2}/2$, $\Delta^*_5\le {\Delta^*_2}/3$, and  $f_5\ge -c_5^-=f_5^*$. 
%$\nabla_y P(x^*,y^*)=B'\price(z^*)=\Delta^*$ and, since 
%$\Delta^{*\prime}y=\price(z^*)'By=\price(z^*)'BHy=\Delta^{*\prime}Hy$, the 
%condition reads $\Delta^{*\prime}(f-f^*)\le0$ for every $f=Hy$ with $y$ 
%in $\mc Y$. Using \eqref{eq:rangeH}
%%Parametrizing the range of $H$ according to \eqref{eq:rangeH} 
%%as 
%%$$f_1=a\,,\quad f_2=-\frac a4-\frac t2-\frac b3\,,\quad f_3=f_4=t\,,\quad f_5=b\,,$$
%we obtain
%$$\ba{rcl}
%\Delta^{*\prime}f&=&\Big(\Delta^*_1-\frac{\Delta^*_2}4\Big)f_1
%+\Big(2\Delta^*_3-\frac{\Delta^*_2}2\Big)f_3
%+\Big(\Delta^*_5-\frac{\Delta^*_2}3\Big)f_5\\[7pt]
%&=&\ds-\frac{15403}{2565}\,a-\frac{1927}{7695}\,b\,,
%\ea$$
%where the coefficient of $t$ vanishes since $\Delta^*_3=\Delta^*_2/4$. 
%As both the remaining coefficients are negative and every feasible flow 
%satisfies $a\ge-c^-_1=0$ and $b\ge-c^-_5=-1$, we conclude that 
%$\Delta^{*\prime}f\le\Delta^{*\prime}f^*$ for every feasible $f$. 
Therefore, the first-order condition $\nabla P(x^*,y^*)'((x,y)-(x^*,y^*))\le0$ is satisfied for every $x$ in $\mc X$ and $y$ in  $\mc Y$. Since $P$ is concave, this implies that $(x^*,y^*)$ maximizes $P$ over the convex set $\mc X\times\mc Y$, hence it is a Nash equilibrium. Moreover, every Nash equilibrium shares the same $x^*$ and $By^*$, hence the same 
consumption vector $z^*$. In particular, 
$z^*_4=-1<0$
at every Nash equilibrium of this NCGSO: this shows that \eqref{z>=0} may fail when $H\ne I$, even though all the remaining 
assumptions of Theorem~\ref{theo:pos} are satisfied (including  \eqref{eq:ass_pos} with  $a=12$). Intuitively, since 
link $1$ is directed, by \eqref{eq:rangeH} moving flow from the cheap 
market $1$ toward market $2$ forces a decrease of $f_5$, i.e., a 
drainage from market $4$, where no producer sells, to market $3$; 
this trade-off remains profitable up to the saturation $f^*_5=-c^-_5$, 
driving the consumption of market $4$ below zero.
\end{example}}
%%%%%%%%%%%% OLD EXAMPLE %%%%%%%%%%%%%%
%	\begin{example}
%	 Consider a network $(A, B, c)$ with $n=3$ producers, $m=3$ markets, $l=3$ links, producer-market adjacency matrix $A=I$, and market-link incidence matrix and capacity vectors $$B=\begin{pmatrix}-1&0&1\\1&-1&0\\0&1&-1\end{pmatrix},\quad c^+=(3,5,5)\,,\;\quad		c^-=(5,0,5)\,,		$$respectively. Let the inverse demand functions be $$		\price_1(s)=12-s\,,\quad		\price_2(s)=12-s/4\,,\quad		\price_3(s)=12-6s\,,		$$ which satisfy \eqref{eq:ass_pos} with $a=12$, and let $$H=I+\1 v',\qquadv=\left(7/3,{11}/3,-7\right)\,,$$ which satisfies \eqref{BH=B} as $BH=BI+(B\1) v'=B$.		Assume that producers are playing actions $x_1=(1,0,0)$, $x_2=(0,1,0)$, and $x_3=(0,0,2)$, respectively, so that $\diag(A'x)=(1,1,2)$. We then find		$$\mc B_0(x)=\{y\in \R^{l}\,\mid \, Hy=(3,0,1)\}\,,	$$		%$$\mc B_0(x)= \{ (3,0, 1) + \alpha \1\,,\; \alpha \in \mathbb{R}\}$$				yielding $By=(-2,3,-1)$ and $z=(-1,4,1)$ for all $y$ in $\mc B_0(x)$.		Note $z_1<0$, showing that \eqref{z>=0} might not hold when $H\neq I$.  By contrast, if $H=I$, the best response is $$\mc B^I_0(x)=\{({65}/{31},0,{58}/{31}) + \alpha\1\,,\; \alpha \in \R\}\,,$$ which yields $z=({24}/{31},{96}/{31},{4}/{31})\ge0$ in agreement with Theorem~\ref{theo:pos}. Intuitively, the SO sends flow from market $3$ (low price) to market $1$, but this induces additional flow from market $1$ to market $2$, leading to negative entries.\end{example}}

\subsection{Comparison with Cournot-Stackelberg games}\label{ss:stack}
\begin{figure}
		\centering
		\includegraphics[width=0.25\textwidth]{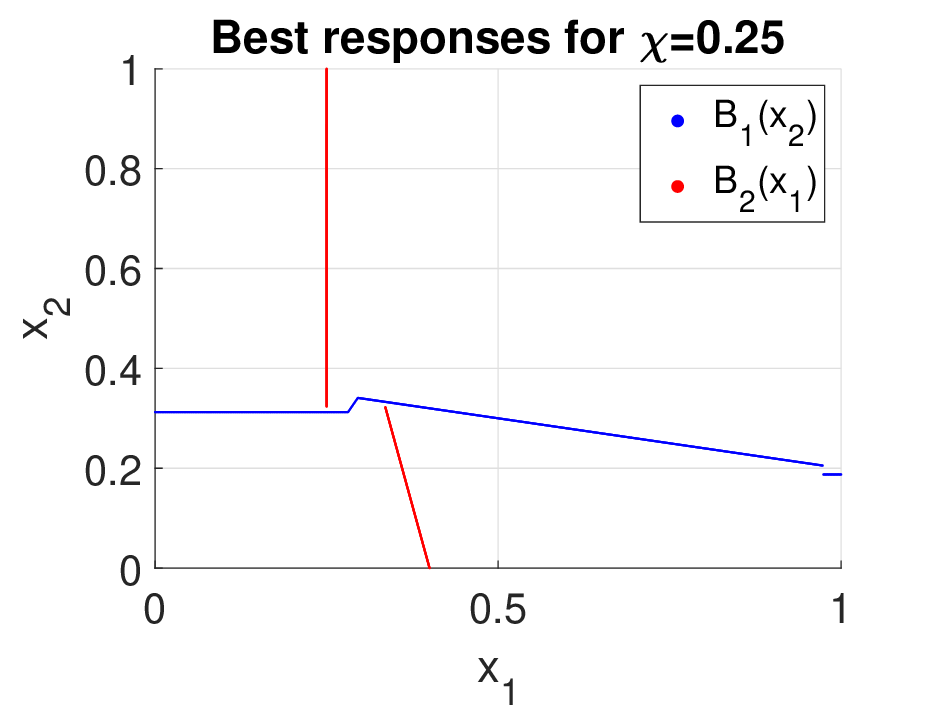}
		\caption{Non-existence of Nash equilibria in the Cournot-Stackelberg game for $\chi={1}/{4}$ (see Example \ref{ex:stack}).}
		\label{fig:stack}
\end{figure}

In our setting, the producers and the SO play simultaneously. By contrast, a significant part of the existing literature \cite{borenstein1997competitive, downward2010cournot} studies a sequential (Stackelberg) game, where producers move first as leaders and the SO responds optimally as a follower. In such models, Nash equilibria might fail to exist, even in very simple settings, e.g., two markets with  identical inverse-demand functions and two symmetric producers, as shown in \cite{borenstein1997competitive}. This failure is structural and  arises from the non-differentiability of producers' effective utility functions once the SO's best response is incorporated. To illustrate this point and to compare equilibrium outcomes when they exist, we analyze below what the equilibrium looks like in the setting of Example \ref{example:basic}(b) when the SO acts as a follower.
	\begin{example}\label{ex:stack}
		Consider two producers having identical cost functions $\cost_i(x_i)=x_i^2$ for $i=1,2$ and utility functions as in \eqref{eq:ut_2p}. In the Cournot-Stackelberg model, after observing $x=(x_1,x_2)$, the SO optimally selects its best response $y=\mc B_0(x)$ as in \eqref{eq:delta_opt}. As a result, Stackelberg equilibria can be characterized as the Nash equilibria of the game between the producers $i=1,2$, having utility functions 
		$$
		\ba{rcl}\bar u_1(x_1,x_2)&=&u_1(x_1,x_2, \mc B_0(x_1,x_2))\\
		&=&x_{1}-2x_{1}^2+x_1\mc B_0(x_1,x_2)\\
		&=&x_{1}-2x_{1}^2+x_1\left(\frac{1+x_{1}-2x_{2}}3\vee(-\chi)\wedge\chi\right)\,,
		\ea$$
		$$
		\ba{rcl}\bar u_2(x_1,x_2)&=&u_2(x_1,x_2, \mc B_0(x_1,x_2))\\
		&=&2x_{2}-3x_{2}^2-2x_2\mc B_0(x_1,x_2)\\
		&=&2x_{2}-3x_{2}^2-2x_2\left(\frac{1+x_{1}-2x_{2}}3\vee(-\chi)\wedge\chi\right)\,.
		\ea$$
		The resulting two-player game is piecewise differentiable, leading to possible discontinuities in the best responses. Specifically, for $\chi\geq 1$, the best response of producer $1$  
		$$
		\mathcal B_1(x_2)
		=
		\left(
		(2 - x_2)/{5}\right)\vee\left(
		2x_2 - 3\chi - 1\right)\wedge\left(
		(1 - \chi)/{4}
		\right)\,,$$ 
is continuous, while, for $0\leq \chi <1$, %it is discontinuous and given by: %and given by
		$$
		\mc B_1(x_2)=
		\begin{cases}\left(\frac{1+\chi }{4}\right) \vee \left(2x_2+3\chi-1\right)\wedge\left(	\frac{2-x_2}{5}\right)&\text{if } x_2\leq x_2^*\\
(1-\chi )/{4}&\text{if } x_2> x_2^*\,.\end{cases}$$
		where $x_2^*=2-\frac{\sqrt{30}}{4}(1-\chi)$, is discontinuous. This discontinuity emerges as the utility of player $1$ admits two coexisting stationary points  $(2-x_2)/{5}$ and $(1-\chi)/4$ whenever $x_2\in \mc I = [\frac{5+11\chi}{8}, \frac{7+15\chi}{11} ]$ (note that this interval is empty for $\chi>1$). As a consequence, the best response is discontinuous at the value $x_2^*$ in $\mc I$ at which player 1 is indifferent between them, i.e., $$\begin{aligned}u_1\left((2-x^*_2)/{5},x_2^*\right)= u_1\left((1-\chi)/{4}, x_2^*\right)\,. \end{aligned}$$
		Similarly, for $\chi \geq 1$, the best response of producer $2$ 
		$$
		\mc B_2(x_1)=\left((2-x_1)/{5}\right)\vee\left(2x_1+3\chi-1\right)\wedge\left((1-\chi)/{3}\right)\,.$$
		is continuous, while it is discontinuous for $0\leq\chi<1$:
		$$\mc B_2(x_1)=\begin{cases}(2-x_1)/{5}&\quad\text{if}\quad 0\leq x_1\leq x_1^*\\(1-\chi)/{3}&\quad\text{if}\quad x_1\geq x_1^*\,,\end{cases}$$
		with $x_1^*=2-\sqrt{5}(1-\chi)\,$. After checking feasibility of all possible combinations of best responses, we find that, for $\chi\geq \chi_+:=1-\sqrt{5}/3$, the Stackelberg-Cournot game admits the unique non-congested Nash equilibrium 
		$$x^*=\left({1}/{3},{1}/{3}\right)\,,\qquad \mc B_0(x^*)=2/{9}\,,$$ while, for $0\leq \chi \leq \chi_-:=(73-24\sqrt{5})/{79}$, it admits the unique congested Nash equilibrium 
$$x^*=\left((1+\chi)/{4}, (1-\chi)/{3}\right)\,,\qquad \mc B_0(x^*)=\chi\,.$$ Finally, for  $\chi_-<\chi<\chi_+$, the game admits \emph{no Nash equilibria}, as shown in Figure \ref{fig:stack} for $\chi={1}/{4}$. 
		
		These findings are consistent with the non-existence results reported in  \cite{borenstein1997competitive,downward2010cournot}, and highlight the fragility of equilibrium existence in sequential formulations. This stands in sharp contrast with the simultaneous-move setting, where existence of pure-strategy Nash equilibria is guaranteed under the conditions in Theorem \ref{theorem:Nash-existence}. By comparing these results with Example \ref{example:basic}, we further find that, when Nash equilibria exist in both settings, they do not always coincide. In particular, in our example, the two Nash equilibria coincide only on the interval $[0,\chi_-]$ (congested equilibria). In the interval $(\chi_-, \chi_+)$, Nash equilibria do not exist in the Stackelberg setting, while a unique congested equilibrium exists in the simultaneous one. Interestingly,  for $\chi_+\leq \chi<{7}/{25}$, the simultaneous game admits a congested equilibrium, while the Stackelberg game admits a non-congested one. Finally,  for $\chi\geq{7}/{25}$, we have a unique non-congested equilibrium in both cases, but these two equilibria do not correspond. \tcb{In the simultaneous case, the price in both markets is $\frac{24}{25}$ %$\price_1(z^*_1)=\price_2(z^*_2)=\frac{24}{25}$ 
			(see \eqref{eq:pr_ex1} where $x_1^*=\frac{8}{25}$  and $x_2^*=\frac{6}{25}$) which is higher than the equilibrium price in the Stackelberg setting, which is  $\price_1(\frac{1}{3}-\frac{2}{9})=\price_2(\frac{1}{3}+\frac{2}{9})=\frac{8} {9}$ with $x_1^*=x_2^*=\frac{1}{3}$. %This might be due to the fact that producers 
			This numerical result is consistent with the intuition that producers may face weaker competitive pressure in simultaneous settings than in settings where flow decisions are determined ex-post (see Remark \ref{rem:stack}).
			%face weaker competitive pressure in simultaneous settings than in settings where flow decisions are determined ex-post. %This suggests that competition may be more effective in practice when net-import decisions at each node are, to a greater extent, determined ex post. 
			Hence, while our theoretical framework guarantees the existence of Nash equilibria, it may overestimate markups and welfare losses.  }%the production quantities $x_1^*=\frac{8}{25}$ and $} %\tcr{Add comment, to be done} %\tcb{Note that in the simultaneous case (Example~\ref{example:basic}b) we have a lower unconstrained aggregate output of $x_1^*+x^*=14/25$ (see \eqref{Nash-sum}), whereas in the sequential case we have a higher aggregate output of $2/3$. This might be due to the fact that when net-import decisions are determined ex post, producers can strategically anticipate how the SO balances flows, which ultimately exerts stronger competitive pressure. Consequently, this comparison highlights that our primary simultaneous model serves as a conservative benchmark that tends to overestimate markups and welfare losses when network capacity is abundant.} 
	\end{example}

%\tcb{As pointed out in Remark  \ref{rem:stack}, producers face weaker competitive pressure in simultaneous settings than in settings where net-import decisions are determined ex post, since marginal changes in a producer’s output at a given node should not affect prices at other nodes. This suggests that competition may be more effective in practice when net-import decisions at each node are, to a greater extent, determined ex post. Hence, while our theoretical model avoids anomalous behaviors, it may overestimate markups and welfare losses. The extent of this potential bias is left for future work.} 

%\tcb{As pointed out in Remark  \ref{rem:stack}, producers have weaker competitive pressure in simultaneous settings than in settings where net-import decisions are determined ex post, as marginal changes in the output of a producer at a specific node should not influence the price at other nodes. This also means that competition would work better in practice where net-import decisions for each node would, to a larger	extent, be determined ex post. Hence, our theoretical model, while avoiding anomalous behaviors, is likely to overestimate mark ups and welfare losses. To which which extent, it is the objective of future work.}
While our results on the existence of Nash equilibria in the simultaneous-move setting, together with related findings in\tcb{\cite{Cai2019}}, stand in sharp contrast to the general non-existence of Nash equilibria in the sequential setting, our results on price differences and link saturation do not rely on any equilibrium notion or on the timing of decisions. Instead, they follow directly from the characterization of the SO’s best response to a given profile of producers’ actions. Therefore, Theorem~\ref{lemma2} and Corollary~\ref{cor:pircemiss}  apply equally to both modeling frameworks and generalize the results in \cite{downward2010cournot} to non-affine demand.

\section{A case study}\label{sec5}
In this section, we  study the prediction accuracy of the proposed model starting from real data from the Italian electricity market \cite{org}. The Italian electricity market, liberalized in 1999, is divided in a spot market ({\it Mercato elettrico a pronti}) and a futures market ({\it Mercato elettrico a termine}). The former is further articulated in a series of branches: the day ahead market ({\it Mercato del giorno prima} --- MGP), several intraday markets ({\it Mercato infragiornaliero} --- MI), the ancillary services market ({\it Mercato del servizio di dispacciamento} --- MSD) and the balancing market ({\it Mercato di Bilanciamento} --- MB). We focus on the day-ahead electricity market, where the majority of electricity is traded. Firms take part into hourly auctions by bidding offers in which they state both the amount they are willing to buy/supply and the corresponding marginal price.  The Italian power system is indeed divided into portions of transmission grids ("zones") with physical limits to transmission of electricity to/from the corresponding neighbouring zones. These transmission limits are determined through a computational model that is based on the balance between electricity generation and consumption. When the market session closes, the GME (energy market manager) computes the aggregate demand and supply and accepts the bids based on economic merit and accounting for the transit limits between zones. The offers for sale are remunerated at the zonal price, i.e., the market-clearing price of the zone they belong to, while the purchase bids are cleared at the National Unique Price ({\it Prezzo Unico Nazionale} --- PUN).

Our analysis is based on two publicly available datasets \cite{org}. The first dataset gathers the zones and the transit limits between them, while the second one contains records of every bid (quantity-price pairs) submitted by the participating firms at each hour of the day. We focus on the following features: the identification code for each producer/consumer participating in the market (\emph{UNIT\_REFERENCE\_NO}), the binary variable determining whether it is a demand or offer bid (\emph{PURPOSE\_CD}), the macro-zone of the bid/offer (\emph{ZONE\_CD}), the quantity offered and the quantity awarded once the market is cleared (\emph{QUANTITY\_NO} and \emph{AWARDED\_QUANTITY\_NO}, respectively), the offered price associated with the transaction and the awarded price (\emph{ENERGY\_PRICE\_NO} and \emph{AWARDED\_PRICE\_NO}, respectively). Quantities are expressed in MWh, marginal prices are expressed in \euro/MWh.

\begin{figure}
	\centering
	\includegraphics[height=5cm]{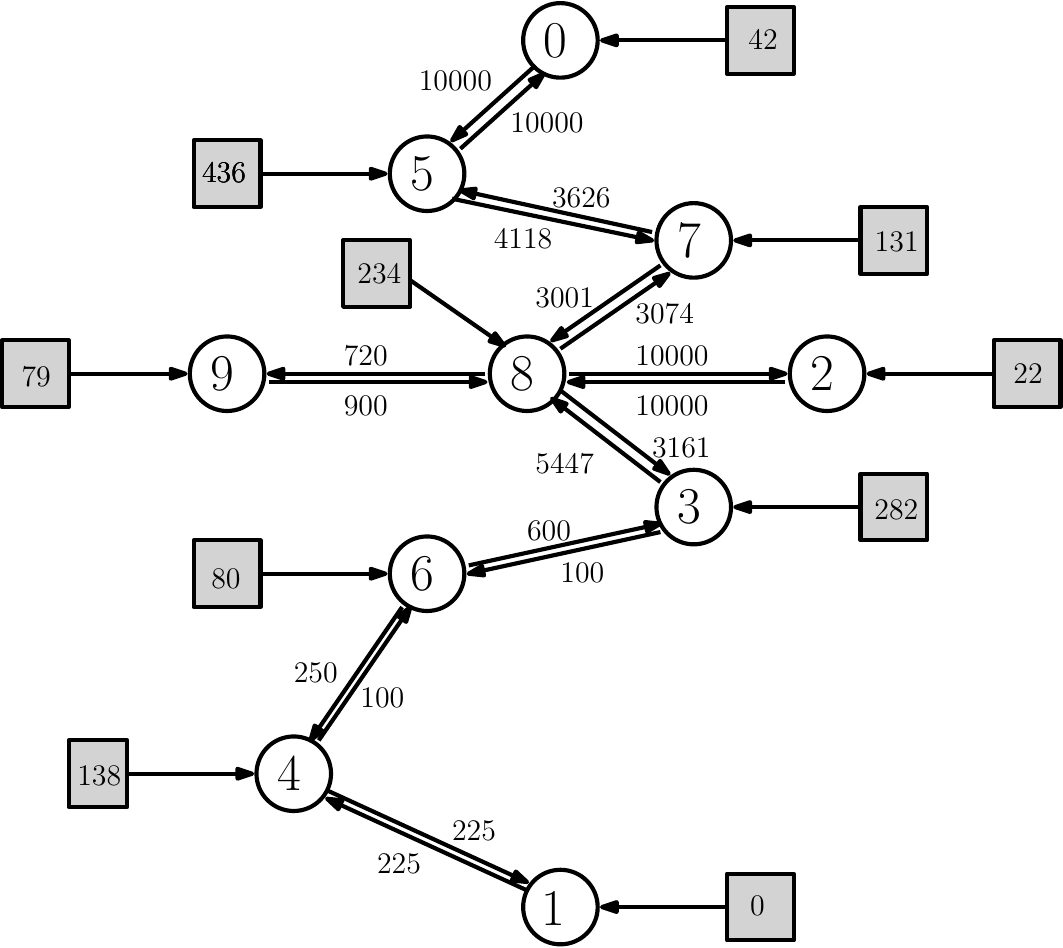}
	\caption{Representation of the Italian power network. \tcb{The number of producers in each market is indicated in the square nodes, while capacities are reported next to the corresponding links.}}
	\label{fig3}
\end{figure}

From the provided data, it is possible to extract a number of statistics describing the composition of the Italian electricity production landscape. In this analysis, we focus on a specific date and time, namely, October 24, 2024,  at 8pm, and we consider only zones where production/consumption takes place and only bidders that are offering positive quantities on such hour of such day. We then obtain the Italian (and neighboring countries) power network shown in Fig. \ref{fig3} consisting of $m=10$ markets (the nodes of the network numbered from 0 to 9) and 9 links. We model the market network accordingly, and we set the capacity constraints $c_+$ and $c_-$ as the publicly available aggregated transit limits across the zones. Note that the network is a tree, therefore the only possible choice for $H$ satisfying \eqref{BH=B} is $H=I$. We have a total of $n=1444$ producers, each selling on a single market. The actual distribution of producers in each market is shown in Table \ref{tab:cs} and in Fig. \ref{fig3} (the red number next to each market) together with the maximum capacity for each line, measured in MWh. 
    
    In the real market, both producers and consumers participate in the auction, submitting quantity/price pairs of bids. However, in order to apply our game-theoretic model, we assume that the only strategic players are the producers and the SO, while demand is treated as non-strategic and is estimated from consumers’ bids. More precisely, we consider affine inverse demand functions  $\price_j(z_j)=\alpha_j-\beta_j z_j$  for all markets $1 \le j \le m$ (measured in \euro/MWh). Producers are assumed to compete only in quantities and to exhibit quadratic cost functions of the form $\cost_i(q_i)=\gamma_i q_i +\costpar_i q_i^2$  for all producers $1 \le i \le n$ (measured in \euro). We set the SO's utility $w$ equal to the Walrasian welfare as in \eqref{Walrasian}, which should be interpreted as a continuous approximation of the bid-based surplus maximization problem solved by the Italian energy market manager, who jointly accepts supply and demand bids in merit order subject to network constraints.

\begin{table}[h]
	\scalebox{0.8}{
		\begin{tabular}{|c|c|c|c|c|c|}
			\hline
			\textbf{Market} & \textbf{Name} & \textbf{$\alpha, \beta$} & \textbf{Num. Prod.} & \textbf{Eq. Price $\frac{\text{\euro}}{\text{MWh}}$} & \textbf{$z^*$}  \\ \hline
			0 & Switzerland & $177.34, \: \:  0.14$ & 42 & 127 & 339.23 \\ \hline
			1 & Malta & $619.49, \: \: 2.63$ & 0 & 197 &  160.20 \\ \hline
			2 & Montenegro & $3933.86, \: \:  9.22$ & 22 & 155 & 409.87 \\ \hline
			3 & South &  $5019.16, \: \:  1.92$ & 282 & 155 & 2537.83 \\ \hline
			4 & Sicily & $5094.16, \: \:  2.03$ & 138 & 197 & 2416.15 \\ \hline
			5 & Nord & $4940.46, \: \:  0.20$ & 436 & 127 & 23390.72 \\ \hline
			6 & Calabria & $5079.5, \: \:  5.97$ & 80 & 197 & 817.72 \\ \hline
			7 & C-Nord & $4747.99, \: \:  1.32$ & 131 & 155 & 3477.99 \\ \hline
			8 & C-Sud & $4818.71, \: \:  0.63$ & 234 & 155 & 7427.78 \\ \hline
			9 & Sardinia & $4918.54, \: \:  4.16$ & 79 & 155 & 1145.22\\ \hline
	\end{tabular}}
	\caption{Parameters, producers, price, and demand in each market at equilibrium.}
	\label{tab:cs}
\end{table}
     
 The value of the parameters has been tuned to fit the real data following similar ideas as in \cite{angelini2023game, vanelli2024game}.  The demand parameters in each zone, that is, $\alpha_j, \beta_j$ for $j$ in $\mc M$, have been estimated in the following way. Starting from all the submitted bids and their merit order, we estimate the demand curve in each zone. We then estimate the demand parameters through a  standard linear regression.  The resulting set of intercepts are presented in Table \ref{tab:cs}. We remark that most of the production takes place in the North (see node 5).  The estimation of the costs parameters  $\gamma_i$ and $\costpar_i$ for each producer $i$ in $\mc N$ is more challenging. The parameter $\gamma_i$ represents the minimum price at which it is strategically convenient for firm $i$ to bid a nonzero quantity to the market. Consequently, we considered the minimum offered marginal price for each production unit, scaled by a factor of $0.99$ in case the bid was not truthful. The parameter $\costpar_i$, instead, was used to incorporate the capacity limits of each producer. More precisely, starting from $\cost'_i(q_i)=\gamma_i +2\costpar_i q_i$, we derive the formula  $ \costpar_i = \frac{\hat{p}-\gamma_i}{2\hat{q}}\,, $ where $\hat{p}$ denotes the awarded price and $\hat{q}$ denotes the awarded quantity. As a consequence, firms that produce higher quantities will result in a lower parameter $\costpar_i$, while $\costpar_i$ is high and the costs grow fast when firms produce low quantities.

\begin{figure}
	\centering
	\includegraphics[height=5cm]{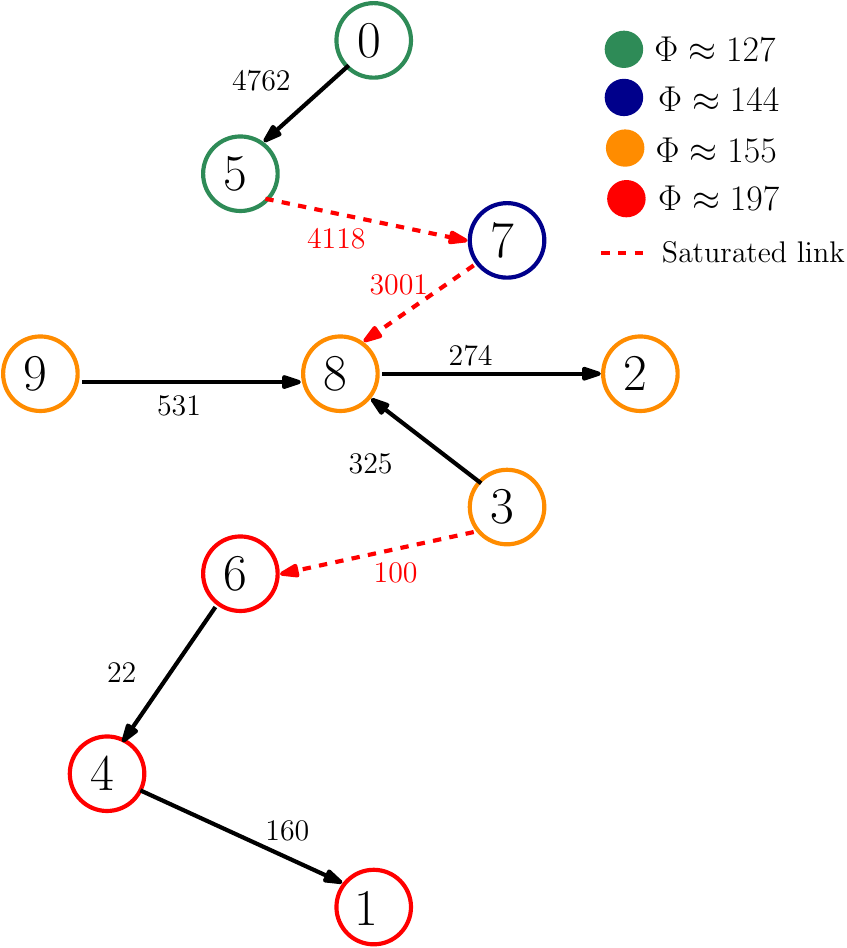}
	\caption{Flows, prices, and capacity bottlenecks in the Italian power network at equilibrium. Different colors highlight groups of markets that have different prices at equilibrium. Dashed lines denote saturated power lines connecting these groups and the arrows indicate the actual direction of the energy flow. Links with negligible flow have been removed. }
	\label{fig4}
\end{figure}

In this setting, the assumptions of Theorem \ref{theorem:potential} are satisfied, so that the NCGSO is an exact potential game with a unique Nash equilibrium that can be found solving \eqref{max-potential}. The resulting equilibrium prices in each market are shown in Figure \ref{fig4} and detailed in Table \ref{tab:cs}.\footnote{The demand at equilibrium in each market is positive, even though the assumption of Theorem \ref{theo:pos} is not satisfied. Our conjecture is that, in this real world example, the demand and the supply in each region are sufficiently balanced to avoid anomalous behaviors.} We observe the emergence of four distinct price groups. The power lines connecting these groups are saturated, facilitating energy flow exclusively from lower-priced markets to higher-priced ones, which is consistent with the results in Corollary \ref{cor:pircemiss}.  Figure \ref{fig4} visually illustrates the groups using distinct colors, with the equilibrium prices labeled next to each node. Dashed lines represent the saturated links, while links with no flow are omitted for clarity. 
\begin{figure}
	\centering
	\includegraphics[width=0.22\textwidth]{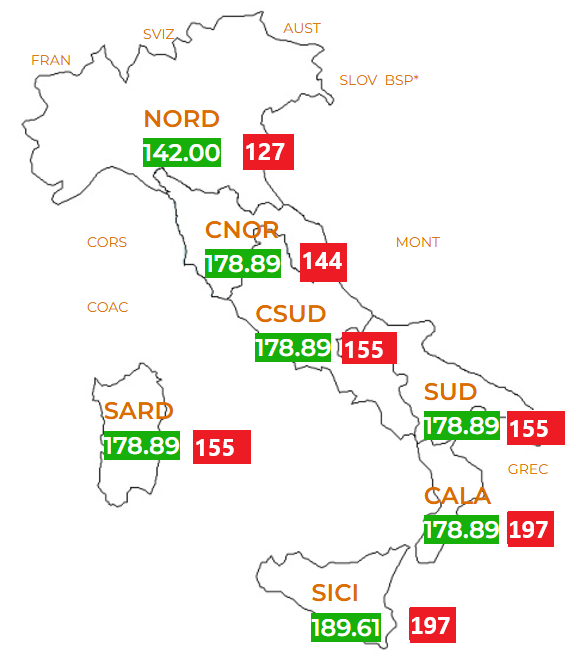}
	\caption{Zonal prices in Italy on October 24th, 2024, at 8pm. This map can be obtained through the \href{https://gme.mercatoelettrico.org/en-us/Home/Results/Electricity/MGP/Results/DemandSupply}{ GME website} by selecting the button MAP. In red, we show for comparison the predicted prices at the Nash equilibrium. }
	\label{fig:map}
\end{figure}
\begin{figure}
	\centering
	\begin{minipage}{0.25\textwidth}
		\centering
		%\hskip -0.3cm
		\includegraphics[width=\linewidth]{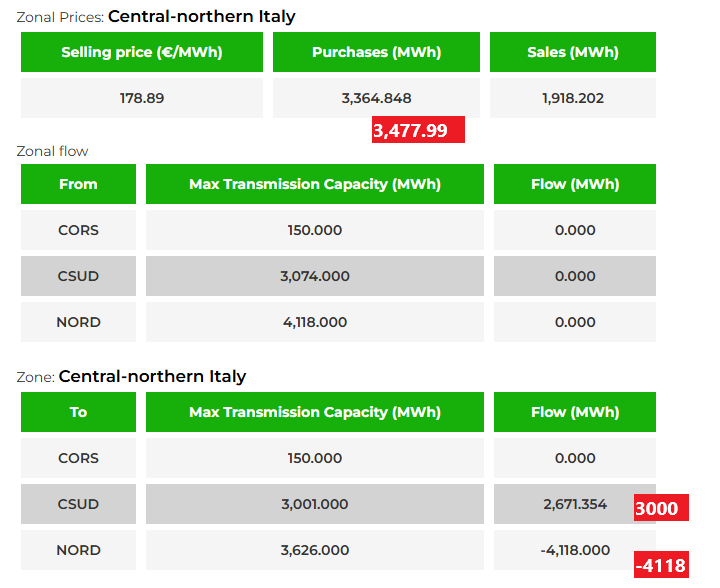}
	\end{minipage}\hfill
	\begin{minipage}{0.15\textwidth}
		\centering
		\hskip -1.3cm
		\scriptsize
		\begin{tabular}{|r|c|c|}
		\hline
		$j$ & %$\text{diag}(A'x^*_\text{eq})_j$
		$Q_j^*$& \textbf{True sales} \\
		\hline
		0 & 5102  & 3239   \\
		\hline
		1 & 0     & 0     \\
		\hline
		2 & 137   & 40    \\
		\hline
		3 & \tcr{2313}  & \tcr{3120}  \\
		\hline
		4 & 2598  & 2210  \\
		\hline
		5 & 22746 & 20954 \\
		\hline
		6 & 696   & 1424  \\
		\hline
		7 & 2361  & 1918  \\
		\hline
		8 & 5553  & 3830  \\
		\hline
		9 & \tcr{616}   & \tcr{1204}  \\
		\hline
		\end{tabular}
%		\captionof{table}{Comparison between equilibrium and actual production.}
	\end{minipage}
	\caption{\tcb{On the left, z}onal prices in Central-Northern Italy on October 24th, 2024, at 8pm. This graph can be obtained through the \href{https://gme.mercatoelettrico.org/en-us/Home/Results/Electricity/MGP/Results/DemandSupply}{GME website} by selecting the button MAP and the zone CNORD. In red, we show for comparison the total quantity purchased in CNORD and the flows at Nash equilibrium. \tcb{On the right, comparison of equilibrium quantities $Q^*=\text{diag}(A'x^*)$ and true sales in each market $j$ in $\mc M$.}}
	\label{fig:lim}
\end{figure}
%\begin{figure}
%	\centering
%	\includegraphics[width=0.35\textwidth]{transit_limits2}
%	\caption{Zonal prices in Central-Northern Italy on October 24th, 2024, at 8pm. This graph can be obtained through the \href{https://gme.mercatoelettrico.org/en-us/Home/Results/Electricity/MGP/Results/DemandSupply}{GME website} by selecting the button MAP and the zone CNORD. In red, we show for comparison the total quantity purchased in CNORD and the flows at Nash equilibrium. }
%	\label{fig:lim}
%\end{figure}
We now compare our findings with the real data from the GME website. In Fig.~\ref{fig:map}, we can see the predicted prices at Nash equilibrium and the real observed ones, respectively. 
Even if we do not recover the exact real prices through our model,  we do, however, predict increasingly higher prices as we go from the north to the south. With our model, we predict three saturated links: from North to Center-North, from Center-North to Center-South, and from Calabria to Sicily. In the real market, only two links are saturated: North to Center-North and South to Calabria. Our model exactly recovers the first one. For the second, saturation is anticipated one step earlier, identifying the South–Calabria link instead of the subsequent one observed in the market. Finally, for the Center-North to Center-South corridor, our model predicts saturation, while the real market shows a flow of 2,671 MWh (versus the 3,000 MWh capacity constraint), as shown in Fig. \ref{fig:lim}. The direction is consistent, and the magnitude is close.
Real data reveal the presence of three distinct price groups. Consistent with the predictions of Corollary \ref{cor:pircemiss}, the links connecting these groups are saturated, facilitating flow from lower-priced markets to higher-priced ones, while no flow occurs in the opposite direction. This behavior is precisely mirrored in our simulations, albeit with the emergence of four price groups.
\tcb{This discrepancy stems from the interplay between modeling choices and parameter estimation. Interestingly, even if simultaneous Cournot models	typically predict lower outputs compared to sequential 
	bid-based ones, Fig. \ref{fig:lim} (right) shows our model generally overestimates production (except in markets $3$ and $9$). The underestimation effect is then outweighed by the lineari demand and quadratic costs assumption, whose parameters are estimated from historical bids (instead of true costs) and calibrated to rise steeply to capture unmodeled capacity constraints. %our quadratic cost assumption with parameters estimated from historical bids rather than true marginal costs and calibrated to rise steeply to absorb capacity constraints that are not modeled explicitly. %Instead of using true marginal costs and explicit capacity limits, we assume a quadratic cost structure and %a quadratic cost specification inferred from historical bids rather than true marginal costs and implicitly absorbing capacity constraints via parameter calibration. 
	Ultimately, these output shifts alter local prices and congestion patterns, directly impacting price zone partitions.  We expect this effect to be amplified when $H \neq I$, as the SO must account for physical flows where a single congested line can induce price separation. Further inaccuracies may arise due to the estimation of $H$ and approximation errors in the linearized DC power flow model.} %Furthermore,  estimating the matrix $H$ compounds approximation errors inherent in the linearized DC power flow model.}

\section{Conclusion} \label{sec6}
We have studied a networked Cournot competition involving producers and a SO competing on multiple markets connected by links with finite capacity. This model is suited to describe energy marketplaces in which links connecting markets represent physical power lines. We have first provided conditions for existence and uniqueness of Nash equilibria. We have then proved a general result concerning the optimal SO's action and the presence of critical links  in the power network where the flow is saturated. %and only moves following a specific direction. 
This result allows us to shed light on the implications of capacity bottlenecks in the power network on the emergence of price differences between different markets. This phenomenon is often observed in the real world  and our model replicates this behavior, as pointed out in the case study of the Italian day-ahead electricity market. 
Future research directions include the design of network intervention policies
%Possible problems involve finding the critical cut and how to optimally create new links or allocate additional capacity among the lines of the power network in order to level price differences or maximize welfare. Future research includes also 
and extensions to supply function equilibria.

%\section*{Acknowledgments}
%The authors gratefully acknowledge Sergio Augusto Angelini for his help in developing the code for the case study.
\bibliographystyle{ieeetr}
\bibliography{bib}

\begin{IEEEbiography}
[{\includegraphics[width=1in,height=1.25in,clip,keepaspectratio]{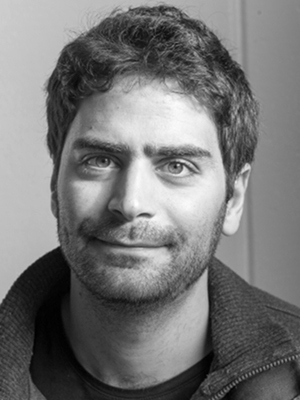}}]
{Giacomo Como} is  a  Professor at  the Department  of  Mathematical  Sciences,  Politecnico di  Torino,  Italy,  and a Senior Lecturer  at  the  Automatic  Control  Department  of  Lund  University,  Sweden.  He  received the B.Sc., M.S., and Ph.D.~degrees in Applied Mathematics  from  Politecnico  di  Torino,  in  2002,  2004, and 2008, respectively. He was a Visiting Assistant in  Research  at  Yale  University  in  2006--2007  and  a Postdoctoral  Associate  at  the  Laboratory  for  Information  and  Decision  Systems,  Massachusetts  Institute of Technology, from 2008 to 2011. 
He is a Senior Editor  of the \textit{IEEE Transactions on Control of Network Systems} and an Associate Editor of \textit{Automatica} and of the  \textit{IEEE Transactions on Automatic Control}. 
He was an  Associate  Editor  of the  \textit{IEEE Transactions on Network Science and Engineering} and of the \textit{IEEE Transactions on Control of Network Systems} and  the chair  of the  {IEEE-CSS  Technical  Committee  on  Networks  and  Communications}. %He was  the  IPC  chair  of  the  IFAC  Workshop  NecSys'15  and  a  semi-plenary speaker  at  the  International  Symposium  MTNS'16.  
He  is  recipient  of  the 2015  George S. ~Axelby  Outstanding Paper Award.  His  research interests  are in  dynamics,  information,  and  control  in  network  systems, with  applications to socio-technical and cyber-physical-human  systems.
\end{IEEEbiography}
\begin{IEEEbiography}
[{\includegraphics[width=1in,height=1.25in,clip,keepaspectratio]{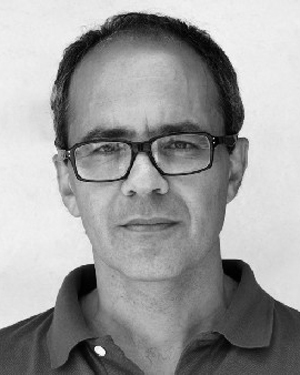}}]
{Fabio Fagnani} received the Laurea degree in Mathematics from the University of Pisa and the Scuola Normale Superiore, in 1986 and the Ph.D.~degree in Mathematics from the University of Groningen,  in 1991. From 1991 to 1998, he was an Assistant Professor of Mathematical Analysis at the Scuola Normale Superiore. In 1997, he was a Visiting Professor at the Massachusetts Institute of Technology (MIT), Cambridge, MA. Since 1998, he has been a Professor of Mathematical Analysis at Politecnico of Torino, where he was the Coordinator of the PhD program in Mathematics for Engineering Sciences from 2006 to 2012 and the Head of the Department of Mathematical Sciences from 2012 to  2019. He was an Associate Editor for the \textit{IEEE Transactions on Automatic Control}, the \textit{IEEE Transactions on Network Science and Engineering}, and the \textit{IEEE Transactions on Control of Network Systems}. His current research topics are on cooperative algorithms and dynamical systems over networks, inferential distributed algorithms, and opinion dynamics.
\end{IEEEbiography}
\begin{IEEEbiography}
[{\includegraphics[width=1in,height=1.25in,clip,keepaspectratio]{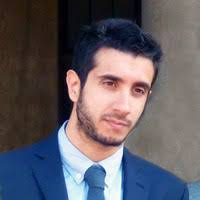}}]
{Leonardo Massai} is a postdoctoral scientist at  the Automatic Control Laboratory, EPFL --- Lausanne, Switzerland working in the DECODE group. 
%on distributed control based on neural networks and machine learning. 
He obtained his B.Sc.~degree in Business Engineering from the University of Pisa in 2012 and his M.Sc.~and Ph.D.~degrees from Politecnico di Torino in 2015 and 2022, respectively. His research interests include optimization, machine learning, and financial mathematics.
\end{IEEEbiography}
\begin{IEEEbiography}
[{\includegraphics[width=1in,height=1.25in,clip,keepaspectratio]{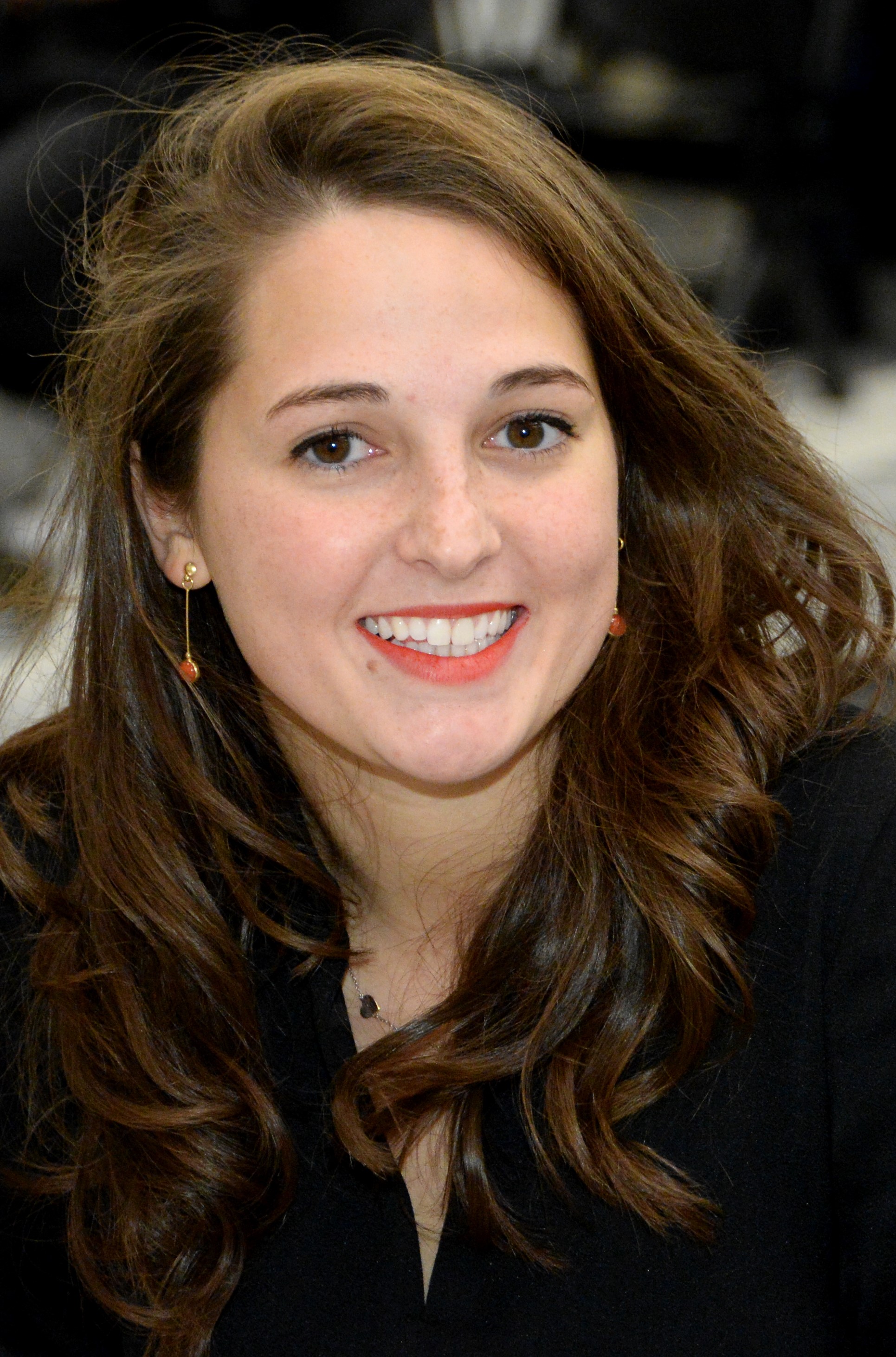}}]
{Martina Vanelli} \tcb{is an Assistant Professor at the Engineering and Technology Institute Groningen (ENTEG), University of Groningen. She} received the B.Sc., M.S., and Ph.D.~degrees in Applied Mathematics  from  Politecnico  di  Torino,  in  2017,  2019, and 2024, respectively. %graduated with honors as a PhD in Applied Mathematics at the Department of Mathematical Sciences of Politecnico di Torino in 2024. She previously received the B.Sc. and the M.S. (\textit{cum laude}) in Applied Mathematics from Politecnico di Torino in 2017 and 2019, respectively. 
\tcb{From 2024 to 2026, she was a Postdoctoral Fellow at the Institute for Information and Communication Technologies, Electronics and Applied Mathematics (ICTEAM), UCLouvain.} %From October 2018 to March 2019, she was a visiting student at Technion, Israel Institute of Technology. 
Her research interests include identification, analysis, and control of multi-agent systems, with application to social and economic networks and power markets. %game theory, and network systems with applications to social and economic networks and power markets.%{Martina Vanelli} received the B.Sc., the M.S. and the Ph.D. degrees in applied mathematics from Politecnico di Torino in 2017, 2019, and 2024, respectively. She is currently a postodoctoral fellow at the ICTEAM laboratory of UCLouvain. From October 2018 to March 2019, she was a visiting student at  Technion, Israel Institute of Technology. Her research interests include game theory, auction theory, and network systems with applications to social and economic networks and power markets.%{Martina Vanelli} is a PhD student in Applied Mathematics at the Department  of  Mathematical  Sciences (DISMA),  Politecnico di  Torino,  Italy. She  received the B.Sc. and M.S. degrees in Applied Mathematics  from  Politecnico  di  Torino,  in  2017 and  2019, respectively. From October 2018 to March 2019, she was a visiting student at  Technion, Israel. Her research interests include game theory, auction theory, and network systems with applications to social and economic networks and power markets.
\end{IEEEbiography}
\vfill

%\bibliography{bib}      

\end{document}